\documentclass[titlepage,12pt]{article}
\usepackage{amsmath,amssymb}
\usepackage[dvips]{graphicx}

\newcommand{\simg}{\ \lower 3pt\hbox{${\buildrel > \over \sim}$}\ }
\newcommand{\siml}{\ \lower 3pt\hbox{${\buildrel < \over \sim}$}\ }

\begin{document}

\renewcommand{\baselinestretch}{1.0}
\begin{center}
\vspace{5cm}
{\LARGE Obliquity evolution of extrasolar terrestrial planets} \\
\vspace{1cm}
{\large Keiko Atobe and Shigeru Ida \\}
\vspace{0.5cm}
E-mail: ida@geo.titech.ac.jp \\
  \textit{Department of Earth and Planetary Sciences, 
  Tokyo Institute of Technology, \\
  2-12-1 Ookayama, Meguro-ku, Tokyo 152-8551, Japan \\}
\vspace{0.5cm}
Icarus, in press 
\end{center}

\vspace{1cm}
%\begin{abstract}

\renewcommand{\baselinestretch}{1.0}
\section*{ABSTRACT}

We have investigated the obliquity evolution of terrestrial planets 
in habitable zones (at $\sim 1$AU) in extrasolar planetary systems,
due to tidal interactions with their satellite and host star
with wide varieties of satellite-to-planet mass ratio 
($m/M_{\rm p}$) and initial obliquity ($\gamma_0$), through numerical 
calculations and analytical arguments. 
The obliquity, the angle between planetary spin axis and
its orbit normal, of a terrestrial planet is one of the key factors 
in determining the planetary surface environments.
A recent scenario of terrestrial planet accretion implies that 
giant impacts of Mars-sized or larger bodies determine
the planetary spin and form satellites.
Since the giant impacts would be isotropic, tilted spins
($\sin \gamma_0 \sim 1$) 
are more likely to be produced 
than straight ones ($\sin \gamma_0 \sim 0$).
The ratio $m/M_{\rm p}$ is dependent on the impact parameters
and impactors' mass.
However, most of previous studies on tidal evolution of the
planet-satellite systems have focused
on a particular case of the Earth-Moon systems 
in which $m/M_{\rm p} \simeq 0.0125$ and $\gamma_0 \sim 10^{\circ}$ 
or the two-body planar problem in which $\gamma_0 = 0^{\circ}$ 
and stellar torque is neglected.
We numerically integrated the evolution of planetary spin
and a satellite orbit with various $m/M_{\rm p}$ (from 0.0025 to 0.05) and 
$\gamma_0$ (from 0$^{\circ}$ to $180^{\circ}$), 
taking into account the stellar 
torques and precessional motions of the spin and the orbit.
We start with the spin axis that almost coincides with 
the satellite orbit normal, assuming that the spin and 
the satellite are formed by one dominant impact.
With initially straight spins, the evolution is
similar to that of the Earth-Moon system.  
The satellite monotonically recedes from the planet 
until synchronous state between
the spin period and the satellite orbital period is realized.
The obliquity gradually increases initially but
it starts decreasing down to zero as approaching the synchronous state.
However, we have found that the evolution with 
initially tiled spins is completely different.
The satellite's orbit migrates outward with almost constant
obliquity until the orbit reaches the critical radius
$\sim 10$--20 planetary radii, but then 
the migration is reversed to inward one.  
At the reversal, the obliquity starts oscillation 
with large amplitude.  The oscillation gradually ceases and
the obliquity is reduced to $\sim 0^{\circ}$ during the
inward migration.
The satellite eventually falls onto the planetary surface or
it is captured at the synchronous state at several planetary radii.
We found that the character change of precession about total
angular momentum vector into that about the planetary orbit normal
is responsible for the oscillation with large amplitude and
the reversal of migration.
With the results of numerical integration and analytical arguments,
we divided the $m/M_{\rm p}$-$\gamma_0$ space into the
regions of the qualitatively different evolution.
The peculiar tidal evolution with initially tiled spins 
give deep insights into dynamics of extrasolar 
planet-satellite systems and
discussions of surface environments of the planets.

%\end{abstract}
Key Words: Extrasolar planets -- Satellites, dynamics --
           Celestial mechanics -- Rotational dynamics --
           Tide, solid body

\section{Introduction} \label{intro}

Recently, extrasolar planets of $5$--$20M_{\oplus}$, which may be
rocky/icy planets, have started being discovered by
development of radial velocity survey 
(e.g., Butler \textit{et al}. 2004, Rivera \textit{et al}. 2005,
Lovis et al. 2006) and gravitational microlensing survey 
(Beaulieu \textit{et al}. 2006), although the
majority of extrasolar planets so far discovered have masses larger than 
Saturn's mass  ($ \gtrsim 100 M_{\oplus}$), which may be gas giant planets.  
If the core accretion model 
(e.g., Mizuno 1980; Bodenheimer and Pollack 1986) is responsible for 
formation of the extrasolar gas giants, 
their high occurrence rate ($\gtrsim 5 \%$) implies
the ubiquity of extrasolar terrestrial planets (e.g., Ida and Lin 2004),
because cores of gas giants are formed through planetesimal 
accretion and failed bodies that are not massive enough for 
gas accretion onto them are no other than terrestrial planets
and icy planets. 
Near-future space telescopes, COROT and KEPLER may find Earth-size planets, 
through transit survey,
including those within so-called "habitable zones," where planets can maintain
liquid water on their surfaces
(e.g., Borucki \textit{et al}. 2003; Ruden 1999).  
%Future missions, 
%TPF and Darwin will be designated to observe atmosphere of the planets
%in habitable zones and detect bio-markers 
%(Beichman \textit{et al}.~1999; Fridlund and Caoaccioni 2002).  

In addition to existence of liquid water,
stable climate on timescales more than $10^9$ yrs may be one of essential
components for planets to be habitable, in particular, for a land-based life. 
Planetary global climate is greatly influenced by insolation distribution
(Milankovitch 1941; Berger 1984,~1989), which is largely
related with obliquity $\gamma$, the angle 
between the spin axis and the orbit normal of the planet (Ward 1974).
For example, if $\gamma > 54^{\circ}$, the planet receives 
more annual-averaged insolation at the poles than at 
the equator and vice versa. 
For high $\gamma$, seasonal cycles at high latitudes would also 
become very pronounced (Williams and Kasting 1997). 
Abe \textit{et al}.~(2005) indicated that obliquity is a very important 
factor for atmospheric transport of water to low-latitude areas. 

Planetary obliquity evolves mainly by tidal interactions with
its satellite and host star.  
All of the terrestrial planets in our solar system do not maintain
their primordial spin state (obliquity $\gamma$ and spin 
frequency $\Omega$). 
Mercury spins with $\gamma = 0$ and $\Omega$ of precisely $3/2$ times 
as large as its orbital mean motion 
(Colombo 1965; Colombo and Shapiro 1965). 
This configuration is an outcome of the tidal interaction with the Sun 
(e.g.,~Goldreich 1966; Goldreich and Peale 1966). 
Venus rotates slowly with a 243-day period and $\gamma \sim 180^{\circ}$.
This spin state could be an equilibrium between gravitational 
and thermal atmospheric tidal torque (Gold and Soter 1969)
or an outcome of friction at a core-mantle boundary (Goldreich and Peale 1970).
Dissipation effects combined with planetary 
perturbations could bring the spin axis to $180^{\circ}$ from 
any initial $\gamma$ (N\'{e}rson de Surgy 1996; Yorder 1997; 
Correia and Laskar 2001). 

The Earth's obliquity 
is gradually increasing with the receding of the Moon as a consequence 
of the tidal dissipation in Earth mainly induced by the Moon 
(e.g., Darwin 1879; Goldreich 1966). 
Figures \ref{fig:EMsys1} show the evolutionary path of the Earth's 
obliquity ($\gamma$) and orbital inclinations of  
the Moon's orbit to the ecliptic ($i$) and 
the Earth's equator ($\epsilon$), which is obtained by 
integrating the present Earth-Moon system back into the past 
(for integration method, see section 2). 
The ranges of oscillation during precession are indicated by shaded 
regions. This plots will be refereed to later.
\begin{center}
[Figure \ref{fig:EMsys1}]
\end{center}

Mars' obliquity would be suffering from a large-scale oscillation 
of $\sim 25^{\circ} \pm 10^{\circ}$ on a timescale $\sim 10^5$--$10^6$ years 
(Ward 1973, 1974, 1979), by a
resonance between a spin precession rate and one of the 
eigenfrequencies of its orbital precession. 
The spin-orbit resonance is a different mechanism to
alter planetary spin state, from the tidal evolution.
Because Earth's spin precession is accelerated out of the resonance
by the Moon, 
Earth's current obliquity fluctuates with 
only $\pm 1.3^{\circ}$ around 
$23.3^{\circ}$ (Ward 1974; Laskar and Robutel 1993; Laskar 1996). 

Here, we focus on 
evolution of planet's spin state and its satellite orbit 
due to the tidal dissipation in the planet caused by the satellite
and the host star.
The spin-orbit resonance in extrasolar planetary systems was addressed 
in detail elsewhere (Atobe \textit{et al}. 2004) and will be 
commented on in section 4.
When the tide raising body (star and/or satellite) orbits around the 
spinning planet, the planet is deformed by the tide
with a certain time interval, resulting in a lag angle 
$\delta$ as illustrated in Fig.~\ref{fig:config1}. 
($\delta$ can be either positive or negative.)
The attraction of the tide raising body
yields a torque on the planet and 
an equal but opposing torque on the body.
In the case of Fig.~\ref{fig:config1}, 
the torque retards the planet spin and
increases the orbital angular momentum and hence semi-major axis of the  
tide raising body. 
If the equatorial plane of the planet does not coincide with the 
orbital plane of the tide raising body, this angular 
momentum exchange also changes the 
obliquity of the planet 
and the orbital inclination of the tide raising body.  
The tidal torque depends on the mass of the tide raising body and strongly 
on the separation between the bodies. The tide raised on Mercury and 
Venus are due to the Sun, while that on Earth is mainly 
induced by the Moon rather than the Sun. 
\begin{center}
[Figure \ref{fig:config1}]
\end{center}

As we will show in later sections, obliquity evolution at $\sim 1$AU
is regulated by a satellite if the satellite-to-planet mass ratio is 
$\gtrsim 0.01$. 
The Moon would have been formed by
a grazing collision with a Mars-mass object during the late 
stage of Earth accretion (e.g., Stevenson 1987, Canup 2004). 
Oligarchic growth model (Kokubo and Ida 1998, 2000) predicts
formation of isolated Mars-mass bodies at $\sim 1$AU in the case of the
minimum mass solar nebula (Hayashi 1981).  
The isolated bodies would start orbit crossing by long term
distant perturbations on timescales longer than than Myrs
(Chambers \textit{et al}. 1996, Iwasaki \textit{et al}. 2002). 
Thus, giant impacts with objects of more than Mars-mass 
would be common and satellites may be produced if the impacts 
are grazing ones.
The satellite mass is determined by total mass 
and angular momentum of the impact-debris disk 
(Ida \textit{et al}.~1997; Kokubo \textit{et al}.~2000), which would be 
regulated by the impact parameter of the collision and the impactor's mass. 

If planetary spin is mostly determined by the giant impact forming a 
satellite (Lissauer \textit{et al}.~2000), 
the spin axis would align with the orbit normal 
of the satellite. Recent N-body simulations of the planet accretion show that 
satellite forming impacts are almost isotropic (Agnor \textit{et al}.~1999; 
Chambers 2001). 
Hence it is expected that the primordial obliquity
$\gamma_0$ has the differential distribution, 
$p(\gamma_0) d \gamma_0 = \frac{1}{2} \sin \gamma_0 d \gamma_0$,
so that the most common initial spins are tilted ones 
($\gamma_0  \sim 90^{\circ}$). 
As we will show, 
the obliquity evolution would be quite different in cases 
of tilted initial spins from the familiar evolution 
in Fig.~\ref{fig:EMsys1}.
In order to investigate obliquity evolution of extrasolar 
terrestrial planets, evolution with wide ranges of initial 
obliquity ($\gamma_0 = 0^{\circ}$--$180^{\circ}$) and satellite mass
should be studied.
Many studies on obliquity evolution have been focusing on
the Earth-Moon system of $\gamma_0 \sim 10^{\circ}$.
Although Counselman (1973) and Ward and Reid (1973) 
addressed more general characteristics and outcome of tidal evolution,
they assumed that the mutual inclinations among 
the satellite's orbit, planet's 
equator and orbit are always zero, i.e., $\gamma = 0$
(for details of Counselman (1973), see section 3.2).
However, as mentioned above, the planetary spin of terrestrial planets 
is more likely to be initially perpendicular to its orbit normal 
($\gamma_0 \sim 90^{\circ}$). 

In this study, we investigate 
the tidal evolution of planet-satellite systems 
with wide ranges of satellite-to-planet mass ratio and initial obliquity,
numerically calculating the planet's spin state and 
satellite's orbit. 
We will show that character change of 
precessional motions during tidal evolution plays an important 
role in producing diversity of the tidal evolution. 
In Section 2, we describe basic equations
for precession and tidal evolution. The initial conditions 
are also described. Section 3 describes the 
results of the simulations and general features of tidal evolution. 
Section 4 and 5 are devoted to discussion and conclusion.

\section{Model and the basic equations} \label{method}

\subsection{Model}

We consider a three-body system composed of a host star with mass $M_{\ast}$, 
a planet with mass $M_{\rm p}$ and physical radius $R_{\rm p}$, 
and a satellite with mass $m$. 
For simplicity, we assume that 
the planet-satellite system is in the circular orbit around the host
star with the mean motion $n_{\rm p}$ and that the satellite is in the 
circular orbit around the planet with the mean motion $n$. 

We adopted the planet-centric frame $(X,Y,Z)$. 
In this frame, the satellite and the 
central star rotate around the planet with $n$ and $n_{\rm p}$, respectively. 
Figure \ref{fig:config2} shows the geometry of the orbital and equatorial 
planes of the planet, and the orbital plane of the satellite. 
$\mathbf{s}$, $\mathbf{k}$, and $\mathbf{n}$ represent
unit vectors in the directions of spin axis of the planet, 
its orbit normal, and the satellite orbit normal, respectively. 
In this frame, $\mathbf{k}$ is a fixed unit vector in the $Z$-direction.
(Although the frame with $\mathbf{n}$ fixed might be better
to discuss the issue of diversity of tidal evolution,
we adopt the frame that is more intuitive.) 
We denote the obliquity of the planet
(the angle between $\mathbf{k}$ and $\mathbf{s}$)
by $\gamma$, the inclination of the satellite orbit to the planet orbit
(the angle between $\mathbf{k}$ and $\mathbf{n}$)
by $i$, and that to the planetary equator 
(the angle between $\mathbf{n}$ and $\mathbf{s}$) by $\epsilon$. 
We assume the planet as an axisymmetric fluid rotating with geometrical 
axis always parallel to its spin angular momentum vector. 
The planetary spin angular velocity is denoted by $\Omega$. 
The satellite spin angular momentum is neglected (Appendix B).
\begin{center}
[Figure \ref{fig:config2}]
\end{center}

\subsection{Basic equations}
\label{subsec:basic_eq}

The scalar angular momentum of the planetary spin and that of the 
satellite orbital motion are denoted by $H$ and $h$. 
Denoting the total precessional torques acting on the planet and the satellite 
by $\mathbf{L}_{\rm p}$ and  $\mathbf{L}_{\rm s}$ and the tidal 
torques by $\mathbf{T}_{\rm p}$ and  $\mathbf{T}_{\rm s}$, 
the equations of motion of the planetary
spin and the satellite orbit are
\begin{eqnarray}
 \frac{d H \mathbf{s}}{dt} 
	&=& \mathbf{L}_{\rm p} + \mathbf{T}_{\rm p}, \label{eq_H}\\
 \frac{d h \mathbf{n}}{dt} 
	&=& \mathbf{L}_{\rm s} + \mathbf{T}_{\rm s}. \label{eq_h}
\end{eqnarray}
The precessional torques 
just rotate the direction of $\mathbf{s}$ and $\mathbf{n}$, 
keeping $H$ and $h$ constant 
(Eqs.~(\ref{eq_Hpre}) and (\ref{eq_hpre}) in Appendix A).
The tidal torques alter $H$ and $h$.
Since angular momentum is also exchanged with 
the planetary orbit, $(H \mathbf{s} + h \mathbf{n})$ does not conserve.
But, since the planet orbital angular momentum is much larger than
$H$ and $h$, we assume that $\mathbf{k}$ is invariant. 

We adopt the method by 
Goldreich (1966) to calculate Eqs.~(\ref{eq_H}) and (\ref{eq_h})
because it follows precessional motions,
although some formalisms (e.g., Mignard 1978, 1979, 1980)
are averaged over the precession.
The precessional motions play a key role in the tidal evolution 
of the system in the case of high initial $\gamma$. 
(For the Earth-Moon system with relatively low initial $\gamma$, 
predicted tidal evolution is hardly affected
by whether precessional motions are taken into account or not.)
Goldreich's method is a multiple averaged ``secular theory,''
utilizing three distinct timescales: orbital periods,
precessional ones, and tidal friction timescale. 
Because tidal evolution timescale is usually much longer than 
precession periods,
$\mathbf{T}_{\rm p}$ and $\mathbf{T}_{\rm s}$ are neglected
in the equations to describe precessional motions.
The equations are
(Goldreich 1966; also see Appendix A)
\begin{eqnarray}
 H \frac{d \mathbf{s}}{dt} 
        & \simeq & \mathbf{L}_{\rm p} 
        = L (\mathbf{s} \cdot \mathbf{n})(\mathbf{s} \times \mathbf{n}) 
	   + K_1 (\mathbf{s} \cdot \mathbf{k})(\mathbf{s} \times \mathbf{k}), 
	    \label{eq_Hpre2} \\
 h \frac{d \mathbf{n}}{dt} 
        & \simeq & \mathbf{L}_{\rm s}  
        = - L (\mathbf{s} \cdot \mathbf{n})(\mathbf{s} \times \mathbf{n}) 
	   + K_2 (\mathbf{n} \cdot \mathbf{k})(\mathbf{n} \times \mathbf{k}),  
	    \label{eq_hpre2}
\end{eqnarray}
where all the quantities are averaged over the orbital periods
(the shortest timescales) of the star and the satellite around the planet.
$H$ and $h$ are constant with time on the precessional periods
(Appendix A).
The first terms in r.~h.~s.~of 
Eqs.~(\ref{eq_Hpre2}) and (\ref{eq_hpre2}) express 
the precessional torques between the planet and satellite that
cause precession around $H \mathbf{s} + h \mathbf{n}$,
and the second ones the stellar torques that cause
the precession around $\mathbf{k}$. 
The constants $L, K_1$, and $K_2$ and detailed expressions of
Eqs.~(\ref{eq_Hpre2}) and (\ref{eq_hpre2}) for numerical integration
are given in Appendix A.
In addition to $H$ and $h$, 
$\Lambda_{Z} = Hx + hy$ (the $Z$-component of total angular momentum in the
planet-satellite system)
and $\chi = K_1 x^2 + K_2 y^2 + L z^2$ (a kind of total potential energy) 
are also conserved (Appendix A), where
$x = \mathbf{s} \cdot \mathbf{k} = \cos \gamma, 
y = \mathbf{n} \cdot \mathbf{k} = \cos i,$ and 
$z = \mathbf{s} \cdot \mathbf{n} = \cos \epsilon. $

The tidal evolution is 
\begin{equation}
 \frac{dH \mathbf{s}}{dt} \simeq \mathbf{T}_{\rm p}; \ \ 
 \frac{dh \mathbf{n}}{dt} \simeq \mathbf{T}_{\rm s}, 
 \label{eq:tide1}
\end{equation}
where $\mathbf{s}$ and $\mathbf{n}$ are averaged over 
precession periods (while those in Eqs.~(\ref{eq_Hpre2}) and
(\ref{eq_hpre2}) are instantaneous ones), and 
$\mathbf{L}_{\rm p}$ and $\mathbf{L}_{\rm s}$ vanish by
the precession averaging. 
$\mathbf{T}_{\rm p}$ and $\mathbf{T}_{\rm s}$ are numerically averaged 
over the precession periods. 
We only consider 
planetary tides induced by the star and the satellite (Appendix B). 
When the tidal torques are included, 
$H$, $h$, $\Lambda_{Z}$, $\chi$, and $a$
that are treated as constant in 
Eqs.~(\ref{eq_Hpre2}) and (\ref{eq_hpre2}) 
change slowly with time (Eqs.~\ref{dH_dh}, \ref{dLamz},
\ref{dKai}, \ref{da2} in Appendix A).
We adopt the constant time lag model by Mignard (1981) and
Touma and Wisdom (1994) for $\mathbf{T}_{\rm p}$ and
$\mathbf{T}_{\rm s}$ (Appendix C).

\subsection{Initial conditions}

If the initial planetary spin state is determined by the satellite forming 
impact, resulting obliquity would have the distribution stated in 
Introduction section and the initial spin period 
would be a few hours if perfect accretion is assumed 
(Agnor \textit{et al}.~1999; Chambers 2001). 
N-body simulations of the accumulation of a satellite in 
a circumplanetary disk predict that the satellite 
is formed in an orbital plane close to the planet's equatorial plane 
($\epsilon \sim 0$) at an orbital radius 2.6--4.6$R_{\rm p}$ 
(Ida \textit{et al}.~1997; Kokubo \textit{et al}.~2000). 
The mass of the formed satellite 
depends on an impact parameter and 
the mass of a projectile and a target. 

We simulated the tidal evolution 
and the evolutionary path of the obliquity and the satellite orbit 
with various $m/M_{\rm p}$ and initial obliquity $\gamma_0$: 
$m/M_{\rm p} = 0.0025 \times j (j=1, 2, \cdots 26)$, 
$\gamma_0 = 10 \times l (l=1, 2, \cdots 8, 10, \cdots 17)$. 
In the numerical calculations, the initial semi-major axis of the satellite,
$a_0$, and the initial rotation period, $D_0$, were chosen to be 
$3.8R_{\rm p}$ and 5 hours. 
Although the simulation is limited to these $a_0$ and $D_0$, 
we will present the dependence of the results on
$a_0$ and $D_0$ through analytical arguments.
As shown later, key quantities to regulate diversity of
tidal evolution are the radius of precession-type change ($a_{\rm crit}$)
and the outer co-rotation radius ($a_{\rm c,out}$).
Although $a_{\rm c,out} \propto D_0^{-2}$ (Eq.~\ref{eq:ac}),
$a_{\rm crit}$ is dependent on $a_0$ and $D_0$ only weakly
($a_0$ does not affect $a_{\rm c,out}$ either).
As long as $D_0$ and $a_0$ differ 
from the above values within a factor of a few,
the overall features of the results presented here do not change. 
Initial inclination of the satellite orbit to 
the planet's equator, $\epsilon_0$, is assumed to $1^{\circ}$. 
We also carried out the calculation with $\epsilon_0 = 5^{\circ}$ and 
$10^{\circ}$ (results are not shown in this paper). 
Increase in $\epsilon_0$ results in an only slight increase 
in the asymptotic value of $i$ at large $a$ and
corresponding oscillation of $\epsilon$ due to precession  
(compare Fig. \ref{fig:EMsys1} with Fig.~\ref{fig:fig1}).
However, other properties of evolution are similar.
If evolution in the case of Fig. \ref{fig:EMsys1} is continued,
$\gamma$ and $\epsilon$ will start decreasing to zero 
as approaching the co-rotation radius.
On the assumption that satellites are formed by a giant 
impact, $\epsilon_0$ would not take larger values. 
Other parameters are assumed that $M_{\ast} = 1M_\odot$, $a_{\rm p} = 1$AU, 
$M_{\rm p}=1M_{\oplus}$, $\alpha=0.33$, $\rho=5.5{\rm gcm}^{-3}$, 
$k_s=0.95$, $k_2=0.30$, and the constant time lag $\delta t =11.5$ min. 
The dependences of the results on these parameters will be
also presented.
We integrate the evolution until the synchronous state is achieved 
or the satellite orbit decays to $2R_{\rm p}$. In the latter case, we 
regard that the satellite falls onto the planet.

\subsection{Precessional motions} 

In order to understand the numerical results of tidal evolution,
we briefly summarize  
characteristic behaviors of precessional motions.
Three different types of precessional motions
are shown in Figs.~\ref{fig:EMsys3}, which were obtained by
numerical integration.
The trajectories of $\mathbf{s}$ and $\mathbf{n}$ 
are projected onto the planetary orbital plane $(X,Y)$, 
in the Earth-Moon system, for 
(a) $a \simeq 5R_{\oplus}$, 
(b) $a \simeq 15R_{\oplus}$, and (c) $a \simeq 60R_{\oplus}$ 
(the present location), respectively.
Although these different precessional motions were already discussed 
in detail by
Goldreich (1966) and Touma \& Wisdom (1994), we present the summary
again, because the character changes in precessional motion
play an important role to produce diversity of obliquity evolution.

When $a \simeq 5R_{\oplus}$, the interaction between the Earth and the 
Moon is much greater than their interaction with the Sun, i.e., 
$L \gg K_1, K_2$ in Eqs.~(\ref{eq_Hpre2}) and (\ref{eq_hpre2}). 
In this case, $\mathbf{s}$ (denoted by a solid curve and 
filled squares) and $\mathbf{n}$ 
(denoted by a dashed curve and filled triangles) 
precess about a common axis with the same precession speeds, 
resulting in constant 
$\epsilon$, the angle between $\mathbf{s}$ and $\mathbf{n}$. 
The common axis precesses slowly about $\mathbf{k}$ ($(X,Y)=(0,0)$) by 
the stellar torques, as indicated in Fig.~\ref{fig:EMsys3}a. 
Hereafter, we call this motion type I precession. 
When $a \simeq 15R_{\oplus}$, torques on the Moon's orbit due to the Earth 
and the Sun are nearly equal, on the other hand, those on the Earth is 
mainly due to the Moon, i.e., $K_2 \sim L > K_1$. 
In this case, $\mathbf{n}$ precesses about 
the average position of $\mathbf{s}$ 
and $\mathbf{k}$, while 
$\mathbf{s}$ tends to precess about $\mathbf{n}$ and is dragged by the 
motion of $\mathbf{n}$, as illustrated in Fig.~\ref{fig:EMsys3}b. 
We call this motion type II precession. 
In the current state, the Sun's torque on the Moon's orbit is much 
larger than the Earth's torque, while the Moon's torque on the Earth is 
about 2 times larger than the Sun's torque, i.e., $K_2 \gg L > K_1$. 
In this case, $\mathbf{n}$ precesses about $\mathbf{k}$ maintaining 
$i$ almost constant. 
Although $\mathbf{s}$ still tends to precesses about $\mathbf{n}$, 
$\mathbf{s}$ actually precesses about $\mathbf{k}$ because 
$\mathbf{n}$ precesses about $\mathbf{k}$ on much shorter period 
(18 years) than the precession period (27000 years) of $\mathbf{s}$ 
about $\mathbf{n}$, 
as illustrated in Fig.~\ref{fig:EMsys3}c. 
We call this motion type III precession. 
\begin{center}
[Figure \ref{fig:EMsys3}]
\end{center}

We define the critical orbital radius of the satellite, 
$a_{\rm crit}$, by 
$K_2 = L$, at which type I precession is transformed into type II. 
From Eqs.~(\ref{eq_Lsp}) and (\ref{eq_Lsc}), 
\begin{equation}
 \frac{K_2}{L} 
	= \frac{3}{2} 
	  \frac{G M_{\ast}}{k_{\rm s} \Omega^2 a_{\rm p}^3} 
	  \left( \frac{a}{R_{\rm p}} \right)^5 
        = \frac{3}{2k_{\rm s}} 
          \left( \frac{n_{\rm p}}{\Omega} \right)^2 
          \left( \frac{a}{R_{\rm p}} \right)^5, \label{K2_L}
\end{equation}
so that $a_{\rm crit}$ is given by 
\begin{equation}
 \frac{a_{\rm crit}}{R_{\rm p}}
	= \left( \frac{2 k_{\rm s}}{3} \right)^{1/5}
            \left( \frac{\Omega}{n_{\rm p}} \right)^{2/5}
	= 18.5 \ k_{\rm s}^{1/5} 
	    \left( \frac{M_{\ast}}{1 M_{\odot}} \right)^{-1/5}
	    \left( \frac{a_{\rm p}}{1 \rm{AU}} \right)^{3/5}
	    \left( \frac{D}{5 \rm{hrs}} \right)^{-2/5}, \label{eq_acrit2}
\end{equation}
where $D$ is the spin period of the planet ($=2 \pi / \Omega$)
at $a_{\rm crit}$.  
For the Earth, $a_{\rm crit} \simeq 17R_{\oplus}$ (Goldreich 1966). 
Although $D$ can take various values according to tidal
evolution from $D_0$, the dependence on $D$ is weak.

Because the ratio $K_2/L$ varies as the fifth power of $(a/R_{\rm p})$, 
the precession of the satellite is almost completely dominated by 
the planetary torque (type I precession) when $a < a_{\rm crit}$ and 
by the stellar torque (type III precession) when $a > a_{\rm crit}$
(Goldreich 1966). 
Type II precession near $a_{\rm crit}$ results in changes 
in all of $\gamma$, $i$, and $\epsilon$.
In the case of initially tilted spins, this fluctuation is so 
large that outward migration of the satellite orbit is 
transformed into inward one (see section 3). 

\section{Results of the tidal evolution} \label{result}

%%%%%%%%%%%%%%%%%%%%%%%%%%%%%%%%%%%%%%%%%%%%%%%%%%%%%%%%%%%%%%%

\subsection{Diversity of tidal evolution}

Our numerical calculations show qualitatively different three types 
(A, B, C) of tidal evolution: 
A) the satellite orbit monotonically expands outward until 
$a$ reaches an outer co-rotation radius,
B) it expands to $\sim a_{\rm crit}$ and 
turns back onto the planet, and 
C) the same as B) but it is locked at a synchronous state 
before falling onto the planet. 
The tidal evolution of the Earth-Moon system belongs to
type A evolution, so that characteristics of 
type A evolution have been well known.  
On the other hand,
type B and C evolution is newly found by us.
The evolution is type B or C if an initial spin is tilted,
as shown below.
Since tilted initial spins are most common, as we stated in 
Introduction section, the newly found type B and C evolution
would be more common than type A for extrasolar terrestrial planets.

\subsubsection{Tidal evolution with initially straight spins}

Figures \ref{fig:fig1} show a typical result of type A evolution with 
$m=0.02M_{\rm p}$ and $\gamma_0 = 10^{\circ}$: 
(a) obliquity $\gamma$, 
(b) $i$, (c) $\epsilon$, and 
(d) the semimajor axis of the satellite ($a$), the estimated inner and 
co-rotation radii ($a_{\rm c,in}$ and $a_{\rm c,out}$) (see below and
section 3.3). 
$\gamma$, $i$ and $\epsilon$ are plotted as functions of $a$. 
The semimajor axis is plotted as a function of time. 
The evolution in this case is qualitatively similar to that of 
the Earth-Moon system in Figs.~\ref{fig:EMsys1}.

In the proximity of the planet, the spin axis of the planet 
($\mathbf{s}$) and the orbit normal of the satellite ($\mathbf{n}$) precess 
about the common axis with nearly constant mutual inclination $\epsilon$ 
(type I precession). 
As long as $a \lesssim a_{\rm crit} \sim 15R_{\rm p}$, 
the satellite orbit migrates outward maintaining 
$\epsilon \sim 0$ (an initial value) and 
accordingly $\gamma \sim i$. 
As the satellite approaches $a_{\rm crit}$, $\mathbf{n}$ tends to precess 
about the average position of $\mathbf{s}$ and $\mathbf{k}$ 
(type II precession). 
Because of this transition of precession, 
$\mathbf{s}$ and $\mathbf{n}$ begin to precess at different rates 
so that $\epsilon$ increases.

When $a > a_{\rm crit}$, $\mathbf{s}$ and $\mathbf{n}$ precess about 
$\mathbf{k}$ with different speeds (type III precession). 
$i$ quickly decreases to zero, while $\gamma$ starts increasing
(Fig.~\ref{fig:fig1}).
These behaviors are explained as follows. 
From Eqs.~(\ref{Hdx}) and (\ref{hdy}) in Appendix A
with tidal torque formula in
Appendix C,
\begin{eqnarray}
\frac{dx}{dt} & = &
- \frac{3}{2} \frac{G m^2 R_{\rm p}^5 k_2 \delta t}{a^6}
        \frac{(y - xz)}{H} (\Omega z - 2n)
- \frac{3}{2} \frac{G M_{\ast}^2 R_{\rm p}^5 k_2 \delta t}{a_{\rm p}^6}
        \frac{(1 - x^2)}{H} (\Omega x - 2n_{\rm p}), \label{dx_sec} \\
 \frac{dy}{dt} & = &
          \frac{3}{2} \frac{G m^2 R_{\rm p}^5 k_2 \delta t}{a^6}
        \frac{(x - yz)}{h} \Omega. \label{dy_sec}
\end{eqnarray}
where $x =\mathbf{s} \cdot \mathbf{k} = \cos \gamma$,
$y=\mathbf{n} \cdot \mathbf{k} = \cos i$,
$z=\mathbf{s} \cdot \mathbf{n} = \cos \epsilon$, and
$\delta t$ is a time lag for distortion of the planet (Appendix C).
In Eq.~(\ref{dx_sec}), the first and the second terms express
the stellar and satellite's tidal torques.
In this region, the satellite tide is dominant.  
Neglecting the second term,
$dx/dt < 0$ if $\Omega z > 2n$, because usually $y > xz$ in this regime.
Since $\Omega \gg n$ except for late phase near the co-rotation radius,
$\gamma$ increases toward
the asymptotic value of $\gamma$,
$\cos^{-1} (2n/ \Omega)$, which is $\sim 90^{\circ}$ for
$\Omega z/n \gg 2$.
Equation (\ref{dy_sec}) includes only a stellar torque.
Since usually $x>yz$ in this regime, 
$dy/dt > 0$ and the stellar torque decreases
$i$ to zero throughout the tidal evolution.

From Eqs.~(\ref{dO}) and (\ref{da2}) with tidal torque formula in
Appendix C, evolution of the spin frequency $\Omega$ and 
the satellite semimajor axis $a$ is
\begin{eqnarray}
 \frac{d \Omega}{dt} &=&
        - \frac{3}{2} \frac{G m^2 R_{\rm p}^5 k_2 \delta t}{a^6}
        \frac{1}{I_{\rm p}} \left\{ \Omega (1+z^2) - 2nz \right\} \nonumber \\
 & & {} \hspace{2cm} - \frac{3}{2}
        \frac{G M_{\ast}^2 R_{\rm p}^5 k_2 \delta t}{a_{\rm p}^6}
        \frac{1}{I_{\rm p}} \left\{ \Omega (1+x^2) - 2n_{\rm p}x \right\},
         \label{dO_sec} \\
 \frac{da}{dt} &=&
          \frac{3}{2} \frac{G m^2 R_{\rm p}^5 k_2 \delta t}{a^6}
        \frac{4a}{h} (\Omega z - n), \label{da_sec}
\end{eqnarray}
where $I_{\rm p}$ is the planet's moment of inertia, given by
$\alpha M_{\rm p} R_{\rm p}^2$ (where $\alpha \leq 2/5$; for the
Earth, $\alpha \simeq 0.33$).
Neglecting the stellar torque (the second term) in Eq.~(\ref{dO_sec})
(assuming $m^2/a^6 \gg M_*^2/a_{\rm p}^6$),
$d \Omega/dt < 0$ if $\Omega > (2 \cos \epsilon/(1+ \cos^2 \epsilon))n$.
Thus, as long as $\Omega > 2 n$, the spin rate $\Omega$ 
decreases for any value of $\epsilon$.
Eventually $\Omega \cos \epsilon $ becomes smaller than $2n$,
so that $dx/dt > 0$ (Eq.~\ref{dx_sec}) and
$\gamma$ begins to decrease, as shown in Fig.~\ref{fig:fig1}c.
The asymptotic value is
$\cos^{-1} (2n/ \Omega) \rightarrow 0^{\circ}$ as
$\Omega z/n \rightarrow 2$.
For $\Omega z/n < 2$, $\gamma$ approaches $0^{\circ}$
regardless of its value.

This decrease in $\gamma$ causes the stable synchronism 
($\Omega \sim n$) with $\gamma \sim i \sim \epsilon \sim 0$. 
Equation (\ref{da_sec}) indicates the satellite is receding from the planet 
when $\Omega z/n > 1$. 
The co-rotation radius $a_{\rm c}$ at which $\Omega z/n=1$ 
depends on the value of $z$. 
There are generally two co-rotation radii in the 
prograde case; the synchronous state is stable at outer one $a_{\rm c,out}$ 
while it is unstable at inner one $a_{\rm c,in}$ (section 3.2). 
At $a > a_{\rm crit}$ in type A evolution here,
$z > 0$, so that $a$ monotonically increases.
As $a \rightarrow a_{\rm c,out}$, the orbital expansion of
the satellite terminates ($da/dt=0$). 
As discussed in the above, 
$\gamma \sim i \sim 0$ (and hence, $\epsilon \sim 0$) 
as approaching $a_{\rm c,out}$,
so that $a_{\rm c,out}$ is given by $\Omega = n$ in this case.

Note that the weak stellar tidal torque 
secularly decreases
the planetary spin and hence $a_{\rm c,out}$ (Eq.~\ref{dO_sec}).
As a result, the satellite orbit slowly decays, being locked
at $a_{\rm c,out}$ (Ward and Reid 1973), 
although the decay timescale is longer than $10^{10}$ years for a planet 
at 1AU (e.g., Goldreich 1966; also see Eq.~\ref{Dt_ast}).

\subsubsection{The cases of initially tilted spins}

\begin{center}
[Figure \ref{fig:fig2}]
\end{center}

Figures \ref{fig:fig2} show the evolution in the case of 
$m=0.01M_{\rm p}$ and $\gamma_0 = 80^{\circ}$. 
During type I precession up to $\sim 18R_{\rm p}$, $\gamma$ and $i$ 
are kept almost constant. 
At $a \sim 18R_{\rm p}$, type II precession starts. 
As a consequence of transition from type I to type II precession, 
$\epsilon$ suddenly starts oscillation with a large amplitude
(Fig.~\ref{fig:fig2}c).
During type I precession, since $\epsilon_0 = 1^{\circ}$ and
$\epsilon$ is almost conserved,
$\mathbf{s}$ and $\mathbf{n}$ always point to
the same direction as illustrated in
the left panel of Fig.~\ref{fig:config3}.  
In the case of type III precession, $\mathbf{s}$ and $\mathbf{n}$ tend to
precess about $\mathbf{k}$ with different frequencies.
If type III precession starts with $\gamma_0$ and $i_0$, 
$\epsilon \sim \epsilon_0 \sim 0$ when $\mathbf{s}$ and 
$\mathbf{n}$ are at the same precession phases, 
while $\epsilon \sim 2 \gamma_0$ when $\mathbf{s}$ and 
$\mathbf{n}$ are at the opposite precession phases
(the right panel of Fig.~\ref{fig:config3}).
When precession is transformed to type II precession
at $a \sim a_{\rm crit}$, $\mathbf{s}$ and $\mathbf{n}$ tend to
precess differently, although not completely.
Hence, when $a$ becomes $\sim a_{\rm crit}$, the precession
tends to have $\epsilon$ 
oscillate in the range up to $\sim 2 \gamma_0$,
which is consistent with the numerical result 
in Fig.~\ref{fig:fig2}c.

\begin{center}
[Figure \ref{fig:config3}]
\end{center}

Equation (\ref{da_sec}) shows that the orbital migration is inward
when $\Omega z/n = \Omega \cos \epsilon/n < 1$.
Because of the large fluctuation of $\epsilon$ caused by large $\gamma_0$, 
$\Omega \cos \epsilon/n < 1$ during large part of precession.
The satellite orbit migrates back and forth during a precession period.
In the case of Fig.~\ref{fig:fig2}, 
the net migration is inward during type II precession.
Approaching the planet, the interaction between the planet 
and satellite dominates the precession again, 
and $\mathbf{s}$ and $\mathbf{n}$ begin to precess about each other 
(back to type I precession). 
The amplitude of precessional oscillation of $\epsilon$
deceases as the satellite orbit comes back to the 
type I precession regime. 
The asymptotic value of $\epsilon$ is larger than $90^{\circ}$
in the case of Fig.~\ref{fig:fig2};
satellite's orbit becomes retrograde around the planet 
(in which $a_{\rm c}$ vanishes).
Hence, satellite eventually falls onto the planet. 

In section \ref{subsubsec:non_coplanar}, 
the reversal of migration at $a \sim a_{\rm crit}$ is
explained in terms of angular momentum and energy
and the condition for the reversal is presented.
In previous studies, only the stellar tidal torque has been
considered for a mechanism for eventual decay of 
initially prograde satellite orbits.
Here we have newly found that planet-satellite tidal interactions
can reverse the orbital migration in the case of 
tilted spins without any help of the stellar tidal torques.

\subsubsection{The cases of initially moderately tilted spins}

Figures \ref{fig:fig3} show the evolution in the case of 
$m=0.04M_{\rm p}$ and $\gamma_0 = 40^{\circ}$. 
In this case, $a_{\rm crit} \sim 14 R_{\rm p}$. 
The evolution is similar to that in 
Fig.~\ref{fig:fig2} until $a$ turns back from $a_{\rm crit}$. 
In this case, however, the satellite's orbital and planetary spin periods 
are locked at stable synchronism. 
Figure \ref{fig:fig3}d shows the time evolution of $a$ and 
the co-rotation radii $a_{\rm c,in}$ and $a_{\rm c,out}$ 
calculated by $L_{Z0} = L_{Z{\rm c}}$ (Eqs.~\ref{eq:L_Z0} and 
\ref{eq:L_Zc}).
Since the co-rotation radii are defined by $\Omega \cos \epsilon = n$, 
it largely oscillates due to the oscillation of $\epsilon$
on the precession period, at $a \sim a_{\rm crit}$. 
In Fig.~\ref{fig:fig3}d, such oscillation occurs from
$0.4 \times 10^5$ years to $2.2 \times 10^5$ years. 
Since the oscillation is so violent, $a_{\rm c,out}$ and $a_{\rm c,in}$ 
during this period is omitted in Fig.~\ref{fig:fig3}d. 
During this period, 
$a_{\rm c,out}$ changes so rapidly that it passes through 
the satellite orbit without capturing the orbit at the synchronism 
($t = 0.5$--$1.0 \times 10^5$ years). 
After the oscillation ceases, 
$a$ approaches $a_{\rm c,out}$ from the outside of 
$a_{\rm c,out}$, and the satellite orbit is captured at $a_{\rm c,out}$. 
The condition for the capture is analytically derived
in section \ref{subsubsec:non_coplanar}. 
This evolution is also newly found one in the present paper.

\begin{center}
[Figure \ref{fig:fig3}]
\end{center}

%%%%%%%%%%%%%%%%%%%%%%%%%%%%%%%%%%%%%%%%%%%%%%%%%%%%%%%%

\subsubsection{High $m$ cases}

If the satellite is massive ($m \gtrsim 0.05M_{\rm p}$), 
the planetary spin and the satellite's orbit become synchronous before 
the occurrence of the precession transition. 
In this case, $\gamma$, $i$, and $\epsilon$ are almost conserved 
as the initial values. We will refer to this type of evolution as 
type A2 evolution and the evolution in section 3.1.1 as A1.

\subsubsection{Small $m$ cases}

If the satellite is light enough, the stellar tidal torques
dominate over the satellite torques.
In this case, $\Omega$ becomes smaller than $n$
before the obliquity becomes zero, then the satellite 
begins to decay toward planet very slowly. 
The subsequent reduction 
of the planetary spin leads to a
synchronous state with planetary mean motion ($\Omega = n_{\rm p}$) with 
$\gamma \sim i \sim \epsilon \sim 0$.  We will refer to this 
as type A3 evolution.

\subsubsection{Retrograde cases}

If an initial planetary spin is determined by a satellite forming 
impact, retrograde spins ($\gamma_0 > 90^{\circ}$) have equal probability 
to prograde spins ($\gamma_0 < 90^{\circ}$). 
Since we assume that 
$\mathbf{s}$ and $\mathbf{n}$ align initially ($\epsilon_0 \sim 0$),
$i_0 > 90^{\circ}$ when $\gamma_0 > 90^{\circ}$.
When the satellite tide is dominant, 
the second terms of r.h.s. of  
Eqs.~(\ref{dx_sec}) and (\ref{dO_sec}) are negligible.
Compared with the prograde cases, $x$ and $y$ change sign
while $z$ has the same sign.
Then evolution of $\Omega$ and $a$ is the same
(Eqs.~\ref{dO_sec} and \ref{da_sec}), while
the equations for $x$ and $y$,
Eqs.~(\ref{dx_sec}) and (\ref{dy_sec}), change sign. 
So, the obliquity evolution is symmetric about $90^{\circ}$.
For $\gamma_0 > 90^{\circ}$, $\gamma$ decreases toward $90^{\circ}$
until $\Omega z/n < 2$.
After that the obliquity increases to $180^{\circ}$. 
In type A3 evolution, however,
the secular change in $x$ is dominated by 
the second term in r.h.s. of Eq.~(\ref{dx_sec}). 
Then, the obliquity eventually decreases toward $0^{\circ}$.

\subsubsection{Parameter dependence}

Figure \ref{fig:numerical} 
shows all the results obtained by numerical calculations. 
Triangles represent A2 evolution. 
Type A1 evolution with $\gamma \rightarrow 0^{\circ}$ 
(for $\gamma_0 < 90^{\circ}$) or $\gamma \rightarrow 180^{\circ}$ 
(for $\gamma_0 > 90^{\circ}$) are plotted by open circles. 
The results with $\gamma \rightarrow 0^{\circ}$ for
$\gamma_0 > 90^{\circ}$ (type A3), are plotted by filled squares. 
For $\gamma_0 < 90^{\circ}$, the boundary between A1 and A3 is not 
clear in the numerical results, so that 
A3 for $\gamma_0 < 90^{\circ}$ is also plotted by open circles. 
Type B and C evolution is expressed by crosses and 
filled circles, respectively. 
The regions of the qualitatively different tidal evolution are 
clearly divided in the $m/M_{\rm p}$--$\gamma_0$ plane. 
In section \ref{subsubsec:non_coplanar}, 
we analytically derive the boundaries of the regions. 

\begin{center}
[Figure \ref{fig:numerical}]
\end{center}

%%%%%%%%%%%%%%%%%%%%%%%%%%%%%%%%%%%%%%%%%%%%%%%%%%%%%%%%%%%%%%%%

\subsection{Angular momentum and energy}

During the tidal evolution of the system, angular momentum ($L$)
is transfered among the planet, the satellite and the host star, 
while the total mechanical energy ($E$) decreases through tidal dissipation
in the planet.  In this subsection, 
we first summarize the arguments in terms of $L$ and $E$ for the planar problem 
developed by Counselman (1972),
and generalize it to non-planar systems 
to explain diversity of tidal evolution found by 
our numerical simulation.

\subsubsection{Tidal evolution of co-planar planet-satellite system}

If the stellar torques are not included and 
$\epsilon_0 =0$, the tidal evolution is a completely planar problem.
Because the stellar tides are neglected, type II and III precessions 
do not exist. Due to $\epsilon = 0$, type I precession does not exist, either.
In the planar case, total angular momentum $L$ and total mechanical energy 
$E$ of the planet-satellite system are given by
\begin{eqnarray}
 L &=& H + h = \alpha M_{\rm p} R_{\rm p}^2 \Omega 
		+ \frac{M_{\rm p} m}{M_{\rm p}+m} na^2, \label{eq:L} \\
 E &=& \frac{1}{2} \alpha M_{\rm p} R_{\rm p}^2 \Omega^2 
		- \frac{GM_{\rm p}m}{2a}. \label{eq:E}
\end{eqnarray}
These equations are deduced to non-dimensional forms, 
\begin{eqnarray} 
 \tilde{L} &=& \tilde{\Omega} + \tilde{n}^{-1}, \label{eq:Ld} \\
 \tilde{E} &=& \tilde{\Omega}^2 - \tilde{n}^2, \label{eq:nd}
\end{eqnarray}
where 
\begin{eqnarray}
 \tilde{\Omega} &=& (\Omega/\sigma)k^{-3}, \label{def_Qd} \\
 \tilde{n}      &=& (n/\sigma)^{1/3} k^{-1}, \label{def_nd} \\
 \tilde{L} &=& L (\alpha M_{\rm p} R_{\rm p}^2 k^3 \sigma )^{-1}, \label{def_Ld} \\
 \tilde{E} &=& E (\alpha M_{\rm p} R_{\rm p}^2 k^6 \sigma^2)^{-1}. \label{def_Ed} 
\end{eqnarray}
In the above,
the frequency $\sigma$ and a parameter $k$ are defined by 
$\sigma = (G M_{\rm p}/R_{\rm p}^3)^{1/2} = (4 \pi G \rho /3)^{1/2}$ and    
$k = (m/\alpha M_{\rm p})^{1/4} (1 + m/M_{\rm p})^{-1/12}$.
The condition of spin-orbit synchronism ($n = \Omega$) is deduced to 
\begin{equation} 
 \tilde{\Omega} = \tilde{n}^3. \label{cond_syn}
\end{equation}
At synchronous points the gradient of $\tilde{E}$ is 
normal to a constant-$\tilde{L}$ contour.

Contours of $\tilde{L}$ and $\tilde{E}$ and the line 
$\tilde{\Omega} = \tilde{n}^3$ are plotted 
by solid, dashed, and dotted curves in the 
$(\tilde{\Omega},\tilde{n})$ plane in Fig.~\ref{fig:cont1}. 
The satellite-planet system evolves 
along a constant-$\tilde{L}$ line
(corresponding to an initial value, $\tilde{L}_0$) 
in the direction of decreasing 
$\tilde{E}$, as indicated by arrows on solid curves.
Note that smaller $|\tilde{n}|$ corresponds to larger $a$.
Evolution with decreasing (increasing) $\vert \tilde{n} \vert$ 
represents orbital expansion (decay) of the satellite.

\begin{center}
[Figure \ref{fig:cont1}]
\end{center}

The lines of constant-$\tilde{L}$, constant-$\tilde{E}$, 
and $\tilde{\Omega}=\tilde{n}^3$ are mutually tangent at 
$(\tilde{\Omega}^{\ast}, \tilde{n}^{\ast})$ and 
$( -\tilde{\Omega}^{\ast}, -\tilde{n}^{\ast})$, where 
$\tilde{\Omega}^{\ast} \simeq 0.439$ and $\tilde{n}^{\ast} \simeq 0.760$.
If $\vert \tilde{L} \vert$ is 
smaller than the critical value 
$\tilde{L}^{\ast} = \tilde{L}(\tilde{\Omega}^{\ast}, \tilde{n}^{\ast}) 
= - \tilde{L}(-\tilde{\Omega}^{\ast}, -\tilde{n}^{\ast}) \simeq 1.755$, 
constant-$\tilde{L}$ line and the synchronism line (Eq.~\ref{cond_syn})
never cross each other, which means that the satellite monotonically migrates 
inward and inevitably falls onto the planet. 
If $\vert \tilde{L} \vert > \tilde{L}^{\ast}$, there are two synchronous 
points for each $\tilde{L}$: 
the outer points ($a = a_{\rm c,out}$) for 
$\vert \tilde{\Omega} \vert < \tilde{\Omega}^{\ast}$ are stable and 
the inner ones ($a=a_{\rm c,in}$) for 
$\vert \tilde{\Omega} \vert > \tilde{\Omega}^{\ast}$ are unstable, 
as shown in Fig.~\ref{fig:cont1}.

\subsubsection{Tidal evolution of non-planar planet-satellite system}
\label{subsubsec:non_coplanar}

In the non-planar cases ($\epsilon \neq 0$), 
the total angular momentum is given by 
\begin{equation}
 \mathbf{L} = \alpha M_{\rm p} \Omega R_{\rm p}^2  \mathbf{s} 
	+ \frac{M_{\rm p}m}{M_{\rm p}+m}na^2 \mathbf{n}. \label{Vec_L}
\end{equation}
Since the stellar torque is considered here, $\mathbf{L}$ changes 
through exchange with orbital motion of the planet.
However, the stellar torque is restricted in the planet's orbital 
plane ($XY$ plane), $L_Z = \mathbf{L} \cdot \mathbf{k}$ is conserved, where 
\begin{equation}
L_Z = \alpha M_{\rm p} \Omega R_{\rm p}^2  \cos \gamma 
	+ \frac{M_{\rm p}m}{M_{\rm p}+m}na^2 \cos i. \label{def_Lz}
\end{equation}
Note that $L_Z \simeq L_0 \cos \gamma_0$, 
since $\gamma \sim i$ initially. 

Figures \ref{fig:L} show the time evolution of 
normalized $\tilde{L}= \vert \mathbf{\tilde{L}} \vert$, 
$\tilde{L}_Z$,  
and $\tilde{E}$ (Eqs.~\ref{def_Ld} and \ref{def_Ed}), 
in the cases of (a) $m=0.02M_{\rm p}$ and 
$\gamma_0=10^{\circ}$, 
(b) $m=0.01M_{\rm p}$ and $\gamma_0=80^{\circ}$ and 
(c) $m=0.04M_{\rm p}$ and $\gamma_0=40^{\circ}$. 
During the evolution, 
$\tilde{E}$ monotonically decreases throughout the evolution. 
On the other hand, $\tilde{L}_Z$ is conserved in all cases, while 
$\tilde{L}_{XY}$ oscillates and secularly decreases 
because of the stellar torque. 
Therefore $\tilde{L}$ asymptotically approaches
$\tilde{L}_Z$ $(\simeq \tilde{L}_0 \cos \gamma_0)$. 
Using this characteristic, type B evolution is explicitly predicted by 
initial parameters $m/M_{\rm p}$ and $\gamma_0$, as below. 

\begin{center}
[Figure \ref{fig:L}]
\end{center}

In Fig.~\ref{fig:cont2}, we plot the numerically obtained time 
evolution of normalized $\tilde{\Omega} \cos \epsilon$ and $\tilde{n}$
(Eqs.~\ref{def_Qd} and \ref{def_nd}), 
in the cases of (a) $m=0.02M_{\rm p}$ and $\gamma_0=10^{\circ}$
(type A evolution in section 3.1.1), 
(b) $m=0.01M_{\rm p}$ and $\gamma_0=80^{\circ}$ (type B in section 3.1.2) and 
(c) $m=0.04M_{\rm p}$ and $\gamma_0=40^{\circ}$ (type C in section 3.1.3). 
In these figures, we plot $\tilde{\Omega} \cos \epsilon$ 
instead of $\tilde{\Omega}$ 
in order to fix the line representing spin-orbit synchronism 
in the non-coplanar case, $\tilde{\Omega} \cos \epsilon = \tilde{n}^3$,
on the plane. 
Contours of $\tilde{L}$ are plotted for $\epsilon = 0$. 
Since $\epsilon$ is not zero generally, 
actual $\tilde{L}$ is different from $\tilde{L}$ at 
instantaneous points of trajectories in the figure except for 
the initial and final stages in which 
$\epsilon \sim 0$. 
However, these contours provide good guide for the evolution of
the trajectories, as shown below. 
With $\epsilon_0 \sim 0$, initial positions are 
in upper right region ($\tilde{n} > 0$, $\tilde{\Omega} >0$) or 
lower left region ($\tilde{n} <0$, $\tilde{\Omega}<0$)
in Fig.~\ref{fig:cont1}. 
Since evolution is symmetric between the two regions unless stellar 
tidal torque is dominant, we discuss the cases starting with 
$\tilde{n}>0$ and $\tilde{\Omega} >0$ below. 

In the early phase where type I precession is dominated, 
$\epsilon \sim 0$ and $\gamma$ and $i$ are conserved. 
As a consequence, in this phase, the conservation of $\tilde{L}_Z$ 
implies that of $\tilde{L}$ and a trajectory gradually moves along 
a constant-$\tilde{L}$ line corresponding to $\tilde{L}_0$.
When the satellite approaches $a_{\rm crit}$, stellar precession 
torque becomes important, then $\epsilon$ begins to oscillates 
on precession periods (type II precession) and $\tilde{L}$ decreases. 
Because the horizontal axis is $\tilde{\Omega} \cos \epsilon$ in 
Fig.~\ref{fig:cont2}, the oscillation is apparent
in the figure.
When $\gamma_0$ is small, 
the satellite keeps receding from the planet and approaches 
$a_{\rm c,out}$ (type A evolution in Sec.~3.1), as shown in 
Fig.~\ref{fig:cont2}a. However, in the high $\gamma_0$ case 
(Fig.~\ref{fig:cont2}b), the oscillation amplitude of $\epsilon$ 
is so large that
the trajectory strides over $a_{\rm c,out}$.
The oscillation timescale is comparable to 
the precession periods of $\mathbf{s}$. 
Because it may be much shorter than tidal evolution timescale, 
the system cannot be captured at the synchronism. 
Once the trajectory strides $a_{\rm c,out}$, orbital migration of the 
satellite is reversed to inward one.
In the $\tilde{\Omega} \cos \epsilon$-$\tilde{n}$ plane
in Figs.~\ref{fig:cont2}, leftward/downward evolution is reversed
to rightward/upward one (see arrows in Fig.~\ref{fig:cont1}). 

Due to damping of $\tilde{L}_{XY}$,
$\tilde{L}$ decreases to $\tilde{L}_Z$
after the large oscillation. 
If $\tilde{L}_Z < \tilde{L_{\ast}} \sim 1.755$, 
$\tilde{L}$ becomes smaller than 
$\tilde{L}_{\ast}$, so that the satellite eventually falls onto the 
planet (type B evolution in Sec.~3.1), as shown in Fig.~\ref{fig:cont2}b. 
If $\gamma_0$ and hence the oscillation amplitude are 
not large enough 
(equivalently, $\tilde{L}_Z \simeq 
\tilde{L}_0 \cos \gamma_0$ is not small enough) 
to enter the regions of $\tilde{L} < \tilde{L}_{\ast}$, 
the satellite meets $a_{\rm c,out}$ again during 
the inward migration (type C evolution in Sec.~3.1), 
as shown in Fig.~\ref{fig:cont2}c. 
Since the oscillation of $\epsilon$ has been damped,
the system is captured at the synchronism. 
Thus, whether type B or C evolution is predicted by initial 
$\tilde{L}_Z$: type B for $\tilde{L}_Z < \tilde{L}_{\ast}$ and 
C for $\tilde{L}_Z > \tilde{L}_{\ast}$. 

\begin{center}
[Figure \ref{fig:cont2}]
\end{center}

%%%%%%%%%%%%%%%%%%%%%%%%%%%%%%%%%%%%%%%%%%%%%%%%%%%%%%

\subsection{Co-rotation radius}

The condition to stride $a_{\rm c,out}$ separates type B and C 
evolution from type A evolution. 
This condition may be roughly obtained by 
$a_{\rm crit} \simg a_{\rm c,out}^{\infty}$, where 
$a_{\rm c,out}^{\infty}$ is the outer co-rotation radius 
evaluated with $\gamma = i = \epsilon = 0$.
As shown in next subsection, this 
condition is in good agreement with the numerical results. 

The expression of $a_{\rm c,out}^{\infty}$ is 
derived by the conservation of $L_Z$.
$L_Z$ at $a_0$ is
\begin{eqnarray}
L_{Z0} &\simeq&
\left( \alpha M_{\rm p} \Omega_0 R_{\rm p}^2 
+ \frac{M_{\rm p}m}{M_{\rm p}+m}n(a_0) {a_0}^2 \right)
\cos \gamma_0 
\equiv 
\alpha M_{\rm p} \Omega_0 R_{\rm p}^2 (1 + f_0) \label{eq:L_Z0}\\
  f_0 & \simeq & 0.2
        \left( \frac{\alpha}{0.33} \right)^{-1} 
        \left( \frac{\rho}{5.5{\rm gcm}^{-3}} \right)^{1/2} 
	\left( \frac{m}{0.01 M_{\rm p}} \right) 
	\left( \frac{D_0}{5 {\rm hrs}} \right) 
	\left( \frac{a_0}{3.8 R_{\rm p}} \right)^{1/2}, \label{def_f}
\end{eqnarray}
where subscript ``0'' represents initial values. 
Using the definition of the co-rotation radius ($a_{\rm c}$),
$\Omega(a_{\rm c}) \cos \epsilon = n(a_{\rm c})$, the angular
momentum at $a_{\rm c}$ is
\begin{equation}
L_{Z{\rm c}} = \alpha M_{\rm p} n(a_{\rm c}) R_{\rm p}^2 \frac{\cos \gamma}{\cos \epsilon} 
+ \frac{M_{\rm p}m}{M_{\rm p}+m}n(a_{\rm c}) {a_{\rm c}}^2 \cos i.
\label{eq:L_Zc}
\end{equation}
Assuming that at $a_{\rm c,out}$, $i \simeq 0$ and 
the satellite orbital angular momentum is dominated,
$L_{Z{\rm c}} \simeq m n(a_{\rm c,out}) {a_{\rm_c,out}}^2 
              \simeq m \sqrt{GM_{\rm p} a_{\rm_c,out}}$.  
From $L_{Z0} = L_{Z{\rm c}}$,
\begin{eqnarray}
\frac{a_{\rm c,out}^{\infty}}{R_{\rm p}} 
&\sim & \alpha^2 \left(\frac{\Omega_0}{\sigma}\right)^2
	(1 + f_0)^2 
       \left( \frac{m}{M_{\rm p}} \right)^{-2} 	
       \cos^2 \gamma_0 \\
 & \simeq & 
     89 \left( \frac{\alpha}{0.33} \right)^2 
 	\left( \frac{\rho}{5.5{\rm gcm}^{-3}} \right)^{-1/3}
	(1 + f_0)^2 
        \left( \frac{m}{0.01 M_{\rm p}} \right)^{-2} 
	\left( \frac{D_0}{5 {\rm hrs}} \right)^{-2} 
	\left( \frac{M_{\rm p}}{M_{\oplus}} \right)^{-2/3} 
	\cos^2 \gamma_0.
   \label{eq:ac}
\end{eqnarray}
Substitution of $m=0.02M_{\rm p}$ and $\gamma_0=10^{\circ}$ yields
$a_{\rm c,out} \sim 42R_{\rm p}$, which is consistent with 
the numerical results in Fig.~\ref{fig:fig1}d.

%%%%%%%%%%%%%%%%%%%%%%%%%%%%%%%%%%%%%%%%%%%%%%%%%%%%%%%%%

\subsection{Classification of tidal evolution}

\begin{center}
[Figure \ref{fig:a1Q12map1}]
\end{center}

Summarizing the above analytical arguments on the tidal evolution, 
the $m/M_{\rm p}$-$\gamma_0$ plane is 
partitioned into three regions labeled A, B and C as shown 
in Fig.~\ref{fig:a1Q12map1}. 
In region A, the satellite monotonically 
recedes from the planet until $a$ reaches $a_{\rm c,out}$. 
With parameters in region A1, $a_{\rm c,out}^{\infty}$ is larger than 
$a_{\rm crit}$, so that synchronous 
state $\Omega=n$ with 
$\gamma \sim i \sim \epsilon \sim 0$ is realized. 
With parameters in region A2, 
synchronous state is achieved before $a$ reaches $\sim a_{\rm crit}$. 
Then, the obliquity and inclinations ($\gamma$, $i$ and $\epsilon$) do not vary
from initial values. 
In this case, since $\gamma_0 \sim \gamma \sim i$, 
$a_{\rm c,out}$ is given by Eq.~(\ref{eq:ac}) 
without the factor $\cos^2 \gamma_0$.
Hence the boundary is independent of $\gamma_0$. 
In region A3, because of small satellite mass, 
the obliquity evolution is dominated by the stellar tidal torque. 
In regions B and C, the satellite's orbit 
expands until $a_{\rm crit}$. At $a_{\rm crit}$, 
$\epsilon$ begins to oscillate by the character change 
of precessional motion so that the satellites begin to migrate 
inward. The satellite stops at $a_{\rm c,out}$ for the
parameters in region C, 
while it falls onto the planet in region B. 

Regions B and C are separated 
by $\tilde{L}_{\rm Z0} = 1.755$, which is represented by the dashed line in 
Fig.~\ref{fig:a1Q12map1}. 
The boundary of A1 from B or C is given by 
$a_{\rm crit} = a_{\rm c,out}^{\infty}$,
represented by the solid line in Fig.~\ref{fig:a1Q12map1}. 
If $m$ exceeds the mass determined by the above 
equation with $\gamma_0=0^{\circ}$, $a_{\rm c,out}^{\infty}$ is 
smaller than $a_{\rm crit}$ for any $\gamma_0$. 
This separates A2 from B or C. 
The stellar tidal effects are stronger than satellite's one
if the timescale ($\Delta t_{\rm A1}$) for orbital 
expansion to $a_{\rm c,out}^{\infty}$ is
longer than the timescale ($\Delta t_{\rm A3}$)
required for the synchronism by the stellar tide 
($\Delta t_{\rm A1}$ and 
$\Delta t_{\rm A3}$ are given in section 3.5).
This separates A3 from A1. 

In Fig.~\ref{fig:a1Q12map1_2}, we compare these analytical boundaries
with the numerical results in Fig.~\ref{fig:numerical}.  
Circles and filled squares 
represent A1 and A3 evolution, respectively. In the numerical 
calculation, A3 evolution with $\gamma_0 < 90^{\circ}$ is 
similar to A1 evolution. 
Triangles, crosses and filled circles represents 
A2, B and C evolution, respectively. 
The analytically estimated boundaries are in good
agreement with the numerical 
results, especially the boundary of A1 from B or C and 
that separating B and C. Since the transition of precession from type I to II 
is not determined exactly by Eq.~(\ref{eq_acrit2}), 
the boundary which separates A2 from B or C is not clear enough in 
the numerical calculation. 

\begin{center}
[Figure \ref{fig:a1Q12map1_2}]
\end{center}

%%%%%%%%%%%%%%%%%%%%%%%%%%%%%%%%%%%%%%%%%%%%%%%%%%%%%%%

\subsection{Variation magnitude and timescales of obliquity evolution}

So far we have been concerned with diversity of tidal evolution and 
intrinsic dynamics that regulates the diversity. 
In this subsection, we discuss ranges of the obliquity changes and their 
timescales that may be important for implications for planetary climate. 
Typical obliquity changes in individual types of 
evolution are illustrated in Fig.~\ref{fig:obliquity}. 
In general, the variation $\Delta \gamma$ in the obliquity 
is large for large $\sin \gamma_0$.  
For B and C evolution, not only the precession-averaged 
$\Delta \gamma$ but also the oscillation amplitude of $\gamma$ during 
precession is large after the satellite's migration turns back.

\begin{center}
[Figure \ref{fig:obliquity}]
\end{center}

The evolution timescales of the obliquity in A1 and A2 are comparable to   
migration timescale to $a_{\rm c,out}$. 
The migration timescale depends on 
the specific dissipation function $Q$, which
is defined as the inverse of the frictional energy dissipation per cycle of 
the tidal oscillation, related to the phase shift $\delta$ as 
$1/Q \sim 2 \delta \sim 2 \delta t (\Omega - n)$
(e.g., Murray and Dermott 1999). 
With $Q$, Eq.~(\ref{da_sec}) is approximately written as 
\begin{equation}
 \frac{da}{dt} = {\rm sign} (\Omega -n) 
	\frac{3k_2}{Q} \frac{m}{M_{\rm p}} 
	\left( \frac{R_{\rm p}}{a} \right)^5 na. \label{da_secQ}
\end{equation}
Integrating this equation from $a_0$ to $a_{\rm c,out}$ is 
\begin{equation}
 \Delta t = 
   \frac{2}{13} \left( \frac{1}{12 \pi G} \right)^{1/2} 
   \frac{Q}{k_2 \rho^{1/2}} \left( \frac{M_{\rm p}}{m} \right) 
   \left\{ \left( \frac{a_{\rm c}}{R_{\rm p}} \right)^{13/2} 
           - \left( \frac{a_0}{R_{\rm p}} \right)^{13/2} \right\}. 
   \label{Dt_A1}
\end{equation}
(The term of $a_0$ is negligible.)
Substituting Eq.~(\ref{eq:ac}) into Eq.~(\ref{Dt_A1}), 
the timescale for A1 evolution is 
\begin{eqnarray}
 \Delta t_{\rm A1} &\sim& 
   2 \times 10^{10}
   \left( \frac{Q}{10} \right) 
   \left( \frac{k_2}{0.3} \right)^{-1} 
   \left( \frac{\alpha}{0.33} \right)^{13} 
   \left( \frac{\rho}{5.5 \rm{gcm}^{-3}} \right)^{-7} \nonumber \\
 & & {} \hspace{2.5cm} \times
   (1 + f_0)^{13} 
   \left( \frac{m}{0.01M_{\rm p}} \right) ^{-14} \cos^{13} \gamma_0 \ \ 
   [\rm{year}]. \label{Dt_A2}
\end{eqnarray}
Timescale for A2 evolution is given by Eq.~(\ref{Dt_A2}) 
with $\gamma_0=0^{\circ}$ for any $\gamma_0$. 
Figure \ref{fig:a1Q12map2_2} shows A1 evolution time calculated by 
Eq.~(\ref{Dt_A1}) with $Q=12$ (current Earth's value) and $k_2=0.30$. 
Note that $\Delta t_{\rm A1}$ strongly 
depends on the mass ratio $m/M_{\rm p}$ and the initial obliquity 
$\gamma_0$. 
Also note that $Q$ could be larger than the current Earth's value
for planets in early stage, so that
the estimate of evolution timescale includes some uncertainty. 

\begin{center}
[Figure \ref{fig:a1Q12map2_2}]
\end{center}

The satellite's orbital evolution of type B and C is expansion to 
$a_{\rm crit}$ followed by inward migration to $a_{\rm c,out}$ or to 
the planetary surface. 
The characteristic evolution timescales are 
given by integrating Eq.~(\ref{da_secQ}) to $a_{\rm crit}$ 
and multiplying it by a factor 2.  Replacing $a_{\rm c}$ by
$a_{\rm crit}$ in Eq.~(\ref{Dt_A1}) and $\times 2$,
and substituting Eq.~(\ref{eq_acrit2}) into it, 
\begin{eqnarray}
 \Delta t_{\rm B,C}
  &\sim&  
  1.5 \times 10^6 
  \left( \frac{Q}{10} \right) 
  \left( \frac{k_2}{0.3} \right)^{-1} 
  \left( \frac{M_{\ast}}{1 M_{\odot}} \right)^{-13/10} 
  \left( \frac{a_{\rm p}}{1 {\rm AU}} \right)^{39/10}  \nonumber \\
  & & {} \hspace{1.5cm} \times 
  k_s^{13/10}
  \left( \frac{\rho}{5.5 {\rm gcm}^{-3}} \right)^{-1/2} 
  \left( \frac{D}{5{\rm hrs}} \right)^{-13/5} 
  \left( \frac{m}{0.01 M_{\rm p}} \right)^{-1} \ \ [{\rm year}]. 
  \label{t_acp3}
\end{eqnarray}

The A3 evolution timescale is the time necessary for 
$\Omega$ to be decreased to orbital mean motion $n_{\rm p}$
by the stellar torque. 
Neglecting the first term in r.h.s. of Eq.~(\ref{dO_sec}),
\begin{equation}
 \frac{d \Omega}{dt} \simeq 
	- \frac{3k_2}{2 \alpha Q} 
	\left( \frac{M_{\ast}}{M_{\rm p}} \right) 
	\left( \frac{R_{\rm p}}{a_{\rm p}} \right)^3 n_{\rm p}^2. 
 \label{dO_secQ}
\end{equation}
Integrating Eq.~(\ref{dO_secQ}) from $\Omega_0$ to 
$n_{\rm p} = (GM_{\ast}/a_{\rm p}^3)^{1/2}$, 
\begin{eqnarray}
 \Delta t_{\rm A3} 
 &=& \frac{4 \pi}{3} \left( \frac{Q}{k_2} \right) 
     \frac{\alpha \rho}{G} \frac{a_{\rm p}^6}{M_{\ast}^2} 
     (\Omega_0 - n_{\rm p}) \label{t_ast1} \\
 &\sim& 3 \times 10^{11} 
   \left( \frac{Q}{10} \right) 
   \left( \frac{k_2}{0.3} \right)^{-1} 
     \left( \frac{M_{\ast}}{1 M_{\odot}} \right)^{-2} 
     \left( \frac{a_{\rm p}}{1 {\rm AU}} \right)^6 \nonumber \\
 & & {} \hspace{3cm} \times 
     \left( \frac{\alpha}{0.33} \right)   
     \left( \frac{\rho}{5.5 {\rm gcm}^{-3}} \right) 
     \left( \frac{D_0}{5{\rm hrs}} \right)^{-1} \ \   [{\rm year}]. 
 \label{Dt_ast}  
\end{eqnarray}

\section{Discussion} \label{discussion}

Here we comment on 
the spin-orbit resonances along the tidal evolution. 
Other planets' gravitational 
perturbations cause precession of the planetary orbit 
about the invariant plane 
of the planetary system. 
When the spin axis precession frequency ($\nu_{\rm s}$) is commensurate with 
one of eigenvalues of the orbital
precession frequency ($\nu_{\rm o}$), the spin axis can 
fluctuate with large amplitudes. Ward (1973) pointed out that Mars' obliquity 
has suffered from a large scale oscillation because of these spin-orbit 
resonances. Laskar and Robutel (1993) showed that in the absence of massive 
satellites all of the terrestrial planets could have experienced large-scale 
obliquity variations. 
Atobe \textit{et al.} (2004) showed that terrestrial planets 
in habitable zones in extrasolar planetary systems with a gas giant(s) 
generally tend to undergo 
the spin-orbit resonance if they do not have satellites. 
They assumed that 
the planets without satellites underwent nearly head-on collisions 
and had relatively slow spin ($D \gtrsim$ 20 hours). 
Due to the relatively large mass of the 
Moon, Earth's spin axis precesses rapidly and avoids the orbital 
precession frequencies of the planets (Ward 1974; Laskar and Robutel 1993). 

As the orbit expands due to the tidal evolution, 
satellite's contribution to the spin axis 
precession gradually decreases. 
Initially, $\nu_{\rm s} \gg \nu_{\rm o}
\sim 10^{-4}$--$10^{-5}$ rad/year, because of small $D$ and $a$. 
Because a satellite was formed, the impact must have 
been grazing one, so that initial $D$ may be $\sim 5$ hours or less.  
Since $D$ and $a$ increase through the 
tidal evolution, $\nu_{\rm s}$ decreases. 
If a satellite is more massive than the lunar mass ($\sim 0.01M_{\oplus}$), 
the expansion of $a$ is limited because of
small $a_{\rm c,out}$ (Eq.~\ref{eq:ac}). 
Then the decrease of $\nu_{\rm s}$ stops before it goes down to 
the level of $\nu_{\rm o}$,
so that the spin-orbit resonance is avoided. 
If a satellite is light enough, the expansion is so slow that
spin-orbit resonance is not realized 
during main-sequence lifetime ($\sim 10^{10}$ years) of 
solar-type stars.
If a satellite has comparable mass to the lunar mass, 
$\nu_{\rm s}$ eventually becomes as small as 
$\nu_{\rm o}$ within $10^{10}$ years through tidal evolution,
and then obliquity may suffer large variation by the spin-orbit resonance. 
Hence, the spin-orbit resonance may
be important for planets without satellites (always)
and for planets with satellites of about the lunar mass
(at some point in tidal evolution within main-sequence lifetime). 

So far, we have been concerned with planets at $\sim 1$AU, that is, planets 
in habitable zones around solar-type stars ($M_{\ast} \sim 1 M_{\odot}$). 
Habitable zones around lower-mass stars (e.g., M dwarfs)
are well inside 1AU, because of the low stellar luminosity.
As shown in Eq.~(\ref{Dt_ast}), small $a_{\rm p}$ leads 
to rapid removal of angular momentum from the planet-satellite system.
Whatever divergent tidal evolution the system has undergone,
the satellite eventually falls to the planet
and the planet spin eventually becomes straight and
synchronous with its orbital motion by the stellar tidal effects
within $10^8$ years. 
Thus, planets in habitable zones have diversity in planet-satellite
configurations during main-sequence phase of host stars, if 
the stars are solar-type stars or more massive stars.  

Final accretion stage of terrestrial planets would be
multiple collisions of protoplanets (see Sec.~1).
The probability of a grazing impact to produce a satellite
may be higher than head-on one.  
Since the spin produced by the impact is more likely to be tilted,
the formed satellite may be perished onto the planetary surface 
through type B tidal evolution on timescales of Myrs 
or trapped near the planetary surface through type C evolution.  
This means that the satellite may orbit near the planetary surface
when a next protoplanet approaches the planet.
Velocity dispersion of protoplanets would be similar to
their surface escape velocity, which is also similar to Keplerian
velocity of the satellite in the proximity of the planetary surface.
Hence, the energy and angular momentum exchange with
the satellite could alter the orbit of 
the approaching protoplanet.  This effect might
alter scattering/collision cross sections of protoplanets.

\section{Conclusion} \label{concl}

We have investigated obliquity evolution of terrestrial planets due to tidal 
interaction with their satellites and host stars by numerical integration 
and analytical arguments, with wide variety of initial conditions
on the basis of recent N-body simulations of planet accretion.
We found three domains in the parameters of satellite-to-planet mass ratio
($m/M_{\rm p}$) and initial obliquity ($\gamma_0$) in which
the evolution is qualitatively different from one another.
 
If a satellite is formed from a debris disk 
created by an off-center giant impact
and the planetary spin is dominated by 
the impact, the planet would have various initial obliquity and 
a satellite with various mass.
And planetary spin axis and 
orbit normal of the satellite are almost aligned ($\epsilon \sim 0$). 
Since N-body simulations show that
the impacts are almost isotropic, an initial planetary spin axis 
is more likely to be tilted ($\sin \gamma_0 \sim 1$)
from the planet orbit normal than to be aligned with it
($\sin \gamma_0 \sim 0$).
In order to study tidal evolution of extrasolar terrestrial planet-
satellite systems, 
we numerically integrated evolution 
with various $m/M_{\rm p}$ and 
various initial $\gamma_0$, focusing on tilted ones.
(We consider the regions at $\sim 1$AU and assume $\epsilon_0 \sim 0$.)
Most of previous studies on tidal evolution, however, have focused
on a particular case of the Earth-Moon systems 
in which $m/M_{\rm p} = 0.0125$ and $\gamma_0 \sim 10^{\circ}$ 
(e.g., Goldreich 1966; Mignard 1978, 1979, 1980; 
Touma and Wisdom 1994) or the 2-body planar problem 
$\gamma = 0^{\circ}$ (e.g., Counselman 1979).

As the previous studies show, the satellite orbit first expands
with the almost constant obliquity ($\gamma$), 
the inclination of the satellite orbit normal to the planet one ($i$),
and that to the planetary spin axis ($\epsilon$) 
until orbital radius of the satellite ($a$) reaches
$\sim a_{\rm crit} \sim 15R_{\rm p}$ (Eq.~\ref{eq_acrit2}) 
at which the precession about total
angular momentum vector is transformed into that about 
the planet orbital normal. 
We have found that this character change in precession causes diversity of 
tidal evolution as follows:
\begin{enumerate}
\item 
At $a \sim a_{\rm crit}$, $\epsilon$ starts oscillation 
between $\sim 0$ and $\sim 2\gamma_0$.
The enhanced $\epsilon$ instantaneously reduces 
outer co-rotation radius ($a_{\rm c,out}$).
For large $\sin \gamma_0$ cases, 
$a_{\rm c,out}$ strides 
across $a$ without capture at the synchronism 
and the outward migration of the satellite orbit is reversed
to inward one.  

\item
The evolution with the migration reversal is further
divided into two types.
The initial $L_Z$ determines whether the turned-back
satellite falls onto the planet (type B evolution) or
it is captured at $a_{\rm c,out}$ that has jumped inward to 
$\sim 5$--$10R_{\rm p}$ (type C evolution).
  
\item	
With the analytical arguments, 
the parameter space of $m/M_{\rm p}$ and $\gamma_0$ is divided into three 
domains (A, B, and C) of qualitatively different tidal evolution
(Fig.~\ref{fig:a1Q12map1}), which is 
in good agreement with the numerical integration
(Fig.~\ref{fig:a1Q12map1_2}).
Type A evolution is further divided into three types:
A1) evolution similar to the Earth-Moon system, 
A2) evolution with the synchronism at $< a_{\rm crit}$
for $m/M_{\rm p} \simg 0.05$, and
A3) evolution dominated by stellar tidal torque
for $m/M_{\rm p} \siml 0.005$.

\item  
In the final state approaching $a_{\rm c,out}$ or the planetary surface,
$\epsilon \sim 0^{\circ}$.
On the other hand, 
$\gamma, i \sim 0^{\circ}$ for
$\gamma_0 < 90^{\circ}$ while 
$\gamma, i \sim 180^{\circ}$ for $\gamma_0 > 90^{\circ}$
except for A3 evolution.
Typical $\gamma$ evolution in each type is
shown in Fig.~\ref{fig:obliquity}. 
The variation timescales are evaluated in Figs.~\ref{fig:a1Q12map2_2}.
The timescales for B and C evolution are relatively
short ($\lesssim 10^6$ years), because the satellite orbit turns back
at relatively small radius $a_{\rm crit}$.
\end{enumerate}

If the formation of a satellite that
is comparable to or larger than
the Moon's mass and 
planetary spin state are determined by a giant impact, 
the tidal evolution of the satellite orbit and the spin
is most likely to be type B/C that 
have been discovered by the present paper. 
(We will elsewhere discuss
the predicted distributions of $m/M_{\rm p}$ and $\gamma_0$.)
To address dynamics of terrestrial planet-satellite systems
in extrasolar planetary systems, it is important to consider 
the large diversity of tidal evolution.
This diversity may be also important to discuss the 
evolution of surface environments of 
extrasolar terrestrial planets and their habitability.

\section*{Acknowledgments}

This work was supported by grant-in-aid, MEXT 16077202. 
We thank anonymous referees for helpful comments to improve
the paper.

\clearpage

\section*{Appendix A: Equations of precessional motions} 
\label{precess_torque}

Here we briefly summarize the precessional torques
and detailed equations of precessional motions
derived by Goldreich (1966).
The torques acting on the planet ($\mathbf{L}_{\rm p}$) is
sum of those from the satellite and the star: 
$\mathbf{L}_{\rm p} = \mathbf{L}_{\rm ps} + \mathbf{L}_{\rm p \ast}$.
Similarly, torques acting on the satellite is 
$\mathbf{L}_{\rm s} = \mathbf{L}_{\rm sp} + \mathbf{L}_{\rm s \ast}$,
where subscripts ``s'', ``p'' and ``$\ast$'' represent
the satellite, the planet and the star respectively.
$\mathbf{L}_{\rm sp}$ averaged over the orbital periods
is given by 
\begin{equation}
 \mathbf{L}_{\rm sp} 
	= -\frac{G M_{\rm p} m}{a^3} J_2 R_{\rm p}^2 
	(\mathbf{s} \cdot \mathbf{n}) (\mathbf{s} \times \mathbf{n}) 
	\equiv - L (\mathbf{s} \cdot \mathbf{n}) 
		(\mathbf{s} \times \mathbf{n}), 
	\label{eq_Lsp}
\end{equation}
where $a$ is the semi-major axis of 
the satellite's orbit, 
$\sigma = (G M_{\rm p}/R_{\rm p}^3)^{1/2}$, 
and $J_2$ is the planetary oblateness coefficient 
given by 
\begin{equation}
 J_2 = \frac{k_{\rm s} R_{\rm p}^3 \Omega^2}{2GM_{\rm p}} 
     = \frac{k_{\rm s}}{2} \frac{\Omega^2}{\sigma^2}, \label{eq_J2}
\end{equation}
($k_{\rm s}$ is the secular Love number).
For the current Earth, $k_{\rm s}=0.95$ as to give the known 
value of $J_2$ (Munk and MacDonald 1960). 
Replacing $m$ and $a$ in Eq.~(\ref{eq_Lsp}) by 
$M_{\ast}$ and $a_{\rm p}$, 
the averaged precessional torque exerted by the planetary 
equatorial bulge on the 
host star ($\mathbf{L}_{\ast \rm{p}}$) is 
\begin{equation}
 \mathbf{L}_{\ast \rm{p}} 
	= -\frac{G M_{\ast} M_{\rm p}}{a_{\rm p}^3} J_2 R_{\rm p}^2 
	(\mathbf{s} \cdot \mathbf{k}) (\mathbf{s} \times \mathbf{k}) 
	\equiv - K_1 (\mathbf{s} \cdot \mathbf{k}) 
		(\mathbf{s} \times \mathbf{k}), 
	\label{eq_Lcp}
\end{equation}
where $a_{\rm p}$ is the semi-major axis of the planet's orbit. 
The averaged torque exerted by the satellite on the star 
($\mathbf{L}_{\ast \rm{s}}$) is 
\begin{equation}
 \mathbf{L}_{\ast \rm{s}} 
 	= - \frac{3}{4} \frac{G M_{\ast} m}{a_{\rm p}^3} a^2
	 (\mathbf{n} \cdot \mathbf{k}) (\mathbf{n} \times \mathbf{k}) 
	\equiv - K_2 (\mathbf{n} \cdot \mathbf{k}) 
		(\mathbf{n} \times \mathbf{k}). 
	\label{eq_Lsc}
\end{equation} 
Since the satellite is regarded as 
a ring with radius $a$ and mass $m$ in the orbit averaging,
$M_{\rm p}R_{\rm p}^2 J_2$ 
in Eq.~(\ref{eq_Lcp}) is replaced by $\sim ma^2$ in Eq.~(\ref{eq_Lsc}). 

Since $\mathbf{L}_{\rm{p} \ast} = - \mathbf{L}_{\ast \rm{p}}$, 
$\mathbf{L}_{\rm ps} = - \mathbf{L}_{\rm sp}$, 
$\mathbf{L}_{\rm{s} \ast} = - \mathbf{L}_{\ast \rm{s}}$, 
\begin{eqnarray}
 \frac{d H \mathbf{s}}{dt} 
	&=& L (\mathbf{s} \cdot \mathbf{n})(\mathbf{s} \times \mathbf{n}) 
	   + K_1 (\mathbf{s} \cdot \mathbf{k})(\mathbf{s} \times \mathbf{k}), 
	    \label{eq_Hpre} \\
 \frac{d h \mathbf{n}}{dt} 
	&=& - L (\mathbf{s} \cdot \mathbf{n})(\mathbf{s} \times \mathbf{n}) 
	   + K_2 (\mathbf{n} \cdot \mathbf{k})(\mathbf{n} \times \mathbf{k}).  
	    \label{eq_hpre}
\end{eqnarray}
Taking the dot products of Eqs.~(\ref{eq_Hpre}) and 
(\ref{eq_hpre}) with $\mathbf{s}$ and $\mathbf{n}$, respectively,
$dH/dt = dh/dt = 0$.
These equations yield Eqs.~(\ref{eq_Hpre2}) and (\ref{eq_hpre2}).

Dotting $\mathbf{k}$ into Eqs.~(\ref{eq_Hpre2}) and (\ref{eq_hpre2}) 
and forming the combination $hH d(\mathbf{s} \cdot \mathbf{n})/dt$,
\begin{eqnarray}
 \frac{dx}{dt} &=& \frac{L}{H}zw, \label{eq_x} \\
 \frac{dy}{dt} &=& -\frac{L}{h}zw, \label{eq_y} \\
 \frac{dz}{dt} &=& \left( \frac{K_2}{h}y - \frac{K_1}{H}x \right) w, 
	           \label{eq_z} 
\end{eqnarray}
where
$x=\mathbf{s} \cdot \mathbf{k} = \cos \gamma$, 
$y=\mathbf{n} \cdot \mathbf{k} = \cos i$, 
$z=\mathbf{s} \cdot \mathbf{n} = \cos \epsilon$ and 
\begin{equation}
w^2 = [(\mathbf{s} \times \mathbf{n}) \cdot \mathbf{k}]^2 = 
1 - x^2 - y^2 - z^2 + 2xyz. 
\label{eq_w2}
\end{equation}
From Eqs.~(\ref{eq_x}), (\ref{eq_y}) and (\ref{eq_z}), 
we obtain the new constants 
\begin{eqnarray}
 \Lambda_Z &=& Hx + hy, \label{def_Lam} \\
 \chi    &=& K_1 x^2 + K_2 y^2 + L z^2. \label{def_Kai} 
\end{eqnarray}
The conservation of $\Lambda_{\rm z}$ arises because the external (stellar) 
torques on the planet-satellite system lie in the planet's orbit. 
$\chi$ is a kind of potential energy.
Differentiating Eq.~(\ref{eq_w2}) with Eq.~(\ref{eq_x}) to (\ref{eq_z}) 
yields 
\begin{equation}
 \frac{dw}{dt} = \frac{L}{H}z(yz-x) - \frac{L}{h}z(xz-y) + 
	\left( \frac{K_2}{h}y - \frac{K_1}{H}x \right) (xy-z). 
		\label{eq_w} 
\end{equation}

The numerical methods are as follows. 
Precessional motions are calculated by
determining initial data for $x$, $y$, 
 and $z$ using Eqs.~(\ref{def_Lam}), (\ref{def_Kai}) and (\ref{eq_w2}),
 and $w$ with $w = 0$ without any loss of generality,
and simultaneously integrating
 $x$ (Eq.~\ref{eq_x}), $z$ (Eq.~\ref{eq_z}), and $w$ 
 (Eq.~\ref{eq_w}) with the conservation 
 law for $\Lambda_{Z}$ (Eq.~\ref{def_Lam}) to eliminate $y$, 
 using a 4-th order Runge-Kutta method. 
Then we average the tidal torques given by 
 Eqs.~(\ref{Tss_s}) to (\ref{Tcs_k}) in Appendix C over one precessional 
 period with the resulting (instantaneous) values of $x$, $y$, $z$, and $w$. 
Tidal evolution is calculated
by integrating the tidal equations given by $dH/dt$ (Eq.~\ref{dH_dh}), 
 $d \Lambda_Z /dt$ (Eq.~\ref{dLamz}), $da/dt$ (Eq.~\ref{da2}), 
 and $d \chi /dt$ (Eq.~\ref{dKai}) with the updated averaged torques 
 and $x$, $y$, $z$ in the previous step.
With the updated $H$, $\Lambda_{Z}$, $a$, and $\chi$, we go back to
the calculation of precessional motions.

This method of integration assumes that the three timescales of orbit, 
precession and tidal evolution are well separated. 
In some cases, the precessional periods can 
be comparable to or longer than the tidal evolution timescale. In the cases, 
we also integrate directly Eqs.~(\ref{eq_H}) and (\ref{eq_h}), 
instead of the hierarchical method.  
We found that the hierarchical method produces the results in 
good agreement with direct method.

\section*{Appendix B. Equations of tidal evolution}

In the planet-satellite-star system, the principal 
tidal change is brought about by the frictionally retarded 
tide on the planet by the satellite, which results 
in a loss of mechanical energy 
from the planet-satellite system and angular momentum transfer 
from the planet's spin to the orbital motion of the satellite. 
Additional smaller change produced by tide due to the host star
results in dissipation of 
energy and angular momentum transfer from the planet-satellite system 
to orbital motions about the star. 
Assuming a phase-locked spin of the satellite on a 
circular orbit, the tide by the planet on the satellite
does not lead to any secular change. 
Thus, we only consider planetary tides induced by
the star and the satellite.

In Appendix B and C, all quantities are precession-averaged ones. 
For simplicity, notations for the averaging are omitted.  
Dotting Eqs.~(\ref{eq:tide1}) by $\mathbf{s}$ and $\mathbf{n}$ yields 
\begin{equation}
 \frac{dH}{dt}
        = \mathbf{T}_{\rm p} \cdot \mathbf{s}, \; 
 \frac{dh}{dt}
        = \mathbf{T}_{\rm s} \cdot \mathbf{n}, \label{dH_dh}
\end{equation}
Dotting Eqs.~(\ref{eq:tide1}) by $\mathbf{k}$ yields 
\begin{eqnarray}
 \frac{dx}{dt} 
	&=& \frac{\mathbf{T}_{\rm p} \cdot \mathbf{k} - 
	    x \mathbf{T}_{\rm p} \cdot \mathbf{s}}{H}, \label{Hdx}\\
 \frac{dy}{dt} 
	&=& \frac{\mathbf{T}_{\rm s}\cdot \mathbf{k} - 
	    y \mathbf{T}_{\rm s} \cdot \mathbf{n}}{h}. \label{hdy}
\end{eqnarray}
From these equations,
\begin{equation}
 \frac{d \Lambda_{Z}}{dt} =
        (\mathbf{T}_{\rm p} + \mathbf{T}_{\rm s}) \cdot \mathbf{k}.
\label{dLamz}
\end{equation}
Differentiating Eq.~(\ref{def_Kai}) yields 
\begin{equation}
 \frac{d \chi}{dt} =
        \frac{2K_1}{H} x \mathbf{T}_{\rm p} \cdot \mathbf{k} +
        \frac{2K_2}{h} y (\mathbf{T}_{\rm s} \cdot \mathbf{k} +
                          y \mathbf{T}_{\rm s} \cdot \mathbf{n}).
\label{dKai}
\end{equation}
From Eq.~(\ref{dH_dh}) with $H=I_{\rm p} \Omega$, 
\begin{equation}
 \frac{d\Omega}{dt} = \frac{\mathbf{T}_{\rm p} \cdot \mathbf{s}}{I_{\rm p}},
 \label{dO}
\end{equation}
where $I_{\rm p}$ is the planet's moment of inertia, given by 
$\alpha M_{\rm p} R_{\rm p}^2$. 
Since $h=mna^2$, Eq.~(\ref{dH_dh}) yields 
\begin{equation}
 \frac{da}{dt} = 
	2a \frac{ \mathbf{T}_{\rm s} \cdot \mathbf{n} }{h}. 
 \label{da2}
\end{equation}

\section*{APPENDIX C: The averaged tidal torques}

In the present paper, we only consider frictional dissipation 
due to tidal deformation of the planet caused by
the satellite and the star.  
The averaged tidal torques on the satellite ($\mathbf{T}_{\rm s}$)
is the sum of the torque by the satellite tide ($\mathbf{T}_{\rm ss}$) 
and that by the stellar tide ($\mathbf{T}_{s \ast}$).
If the spin axis is not 
aligned with the orbit normal of the tide raising body, 
the spin can carry the tidal bulge 
out of the orbital plane. This out-of-plane bulge 
can produce in-plane torques on a third body. 
Similarly, the torques on the star ($\mathbf{T}_{\rm \ast}$)
is the sum of $\mathbf{T}_{\ast \ast}$ (due to the stellar tide) 
and $\mathbf{T}_{\ast \rm{s}}$ (due to the satellite tide).
The torques on the planet are the opposite of the torques on the exterior 
bodies, $\mathbf{T}_{\rm p} = - (\mathbf{T}_{\ast} + \mathbf{T}_{s}) =
-(\mathbf{T}_{\ast \ast} + \mathbf{T}_{\ast s} + 
\mathbf{T}_{\rm{s} \ast} + \mathbf{T}_{\rm{s} \rm{s}})$.

Adopting the constant time lag model by Mignard (1981) and 
Touma and Wisdom (1994), in which the distortion of the planet is
delayed from the tide raising potential by the time lag $\delta t$,
the torques are
\begin{eqnarray}
 \mathbf{T}_{\rm ss} \cdot \mathbf{s} &=&
	\delta t \frac{k_2 G m^2 R_{\rm p}^5}{a^6}
	\left[ \frac{3}{2} \Omega \sin^2 \epsilon + 
		3 \cos \epsilon (\Omega \cos \epsilon - n) \right], 
 	\label{Tss_s} \\
 \mathbf{T}_{\rm ss} \cdot \mathbf{n} &=&
	\delta t \frac{k_2 G m^2 R_{\rm p}^5}{a^6}
	\left[ 3 (\Omega \cos \epsilon - n) \right], 
	\label{Tss_n} \\
 \mathbf{T}_{\rm ss} \cdot \mathbf{k} &=&
	\delta t \frac{k_2 G m^2 R_{\rm p}^5}{a^6}
	\left[ \frac{3}{2} \Omega ( \cos \gamma - \cos i \cos \epsilon ) 
		+ 3 \cos i ( \Omega \cos \epsilon - n ) \right],
	\label{Tss_k}
\end{eqnarray}
where $k_2$ is the dimensionless tidal Love number 
(Munk and MacDonald 1960), and 
\begin{eqnarray}
 \mathbf{T}_{\ast \ast} \cdot \mathbf{s} &=&
	\delta t \frac{k_2 G M_{\ast}^2 R_{\rm p}^5}{a_p^6}
	\left[ \frac{3}{2} \Omega \sin^2 \gamma + 
		3 \cos \gamma (\Omega \cos \gamma - n_{\rm p}) \right], 
 	\label{Tcc_s} \\
 \mathbf{T}_{\ast \ast} \cdot \mathbf{n} &=&
	\delta t \frac{k_2 G M_{\ast}^2 R_{\rm p}^5}{a_p^6}
	\left[ \frac{3}{2} \Omega (\cos \epsilon - \cos \gamma \cos i)
		+ 3 \cos i (\Omega \cos \gamma - n_{\rm p}) \right],
	\label{Tcc_n} \\
 \mathbf{T}_{\ast \ast} \cdot \mathbf{k} &=&
	\delta t \frac{k_2 G M_{\ast}^2 R_{\rm p}^5}{a_p^6}
	\left[ 3 ( \Omega \cos \gamma - n_{\rm p}) \right].
	\label{Tcc_k}
\end{eqnarray}
Denoting the longitude of the ascending node of the planet's 
equatorial plane on its orbital plane and 
that of the satellite's orbital plane on the planet's 
orbital plane as $\Phi$ and $\Psi$,
\begin{eqnarray}
 \mathbf{T}_{\rm{s} \ast} \cdot \mathbf{s} & = &
	\Omega \delta t \frac{k_2 G m M_{\ast} R_{\rm p}^5}{a^3 a_p^3} 
 	\left[ \frac{3}{8} \sin^2 i \sin^2 \gamma \cos 2(\Phi - \Psi) 
		- \frac{9}{8} \sin^2 i \sin^2 \gamma \right. \nonumber \\
 & & {}	\left. - \frac{3}{4} \cos \gamma \sin \gamma \cos i \sin i 
		\cos (\Phi - \Psi) + \frac{3}{4} \sin^2 \gamma \right], 
 	\label{Tsc_s} \\
 \mathbf{T}_{\rm{s} \ast} \cdot \mathbf{n} & = & 0  \label{Tsc_n} \\
 \mathbf{T}_{\rm{s} \ast} \cdot \mathbf{k} & = &
	\Omega \delta t \frac{k_2 G m M_{\ast} R_{\rm p}^5}{a^3 a_p^3} 
	\left[ - \frac{3}{4} \cos i \sin i \sin \gamma 
		\cos (\Phi - \Psi) \right], 
	\label{Tsc_k}
\end{eqnarray}
and 
\begin{eqnarray}
 \mathbf{T}_{\ast \rm{s}} \cdot \mathbf{s} & = &
	\Omega \delta t \frac{k_2 G m M_{\ast} R_{\rm p}^5}{a^3 a_p^3} 
 	\left[ \frac{3}{8} \sin^2 i \sin^2 \gamma \cos 2(\Phi - \Psi) 
		- \frac{9}{8} \sin^2 i \sin^2 \gamma \right. \nonumber \\
 & & {}	\left. - \frac{3}{4} \cos \gamma \sin \gamma \cos i \sin i 
		\cos (\Phi - \Psi) + \frac{3}{4} \sin^2 \gamma \right], 
 	\label{Tcs_s} \\
 \mathbf{T}_{\ast \rm{s}} \cdot \mathbf{n} & = &
	\Omega \delta t \frac{k_2 G m M_{\ast} R_{\rm p}^5}{a^3 a_p^3} 
	\left[ \frac{3}{4} \cos (\Phi - \Psi) \sin i \sin \gamma \cos^2 i 
		- \frac{3}{4} \cos \gamma \cos i \sin^2 i \right], 
	\label{Tcs_n} \\
 \mathbf{T}_{\ast \rm{s}} \cdot \mathbf{k} & = & 0.  \label{Tcs_k}
\end{eqnarray}

\clearpage

\setlength{\baselineskip}{3ex}

\clearpage
\setlength{\baselineskip}{4ex}

{\Large {\bf Figure captions}} 

\vspace{1em}

\begin{figure}
 \centering 
 \includegraphics[width=10cm,clip]{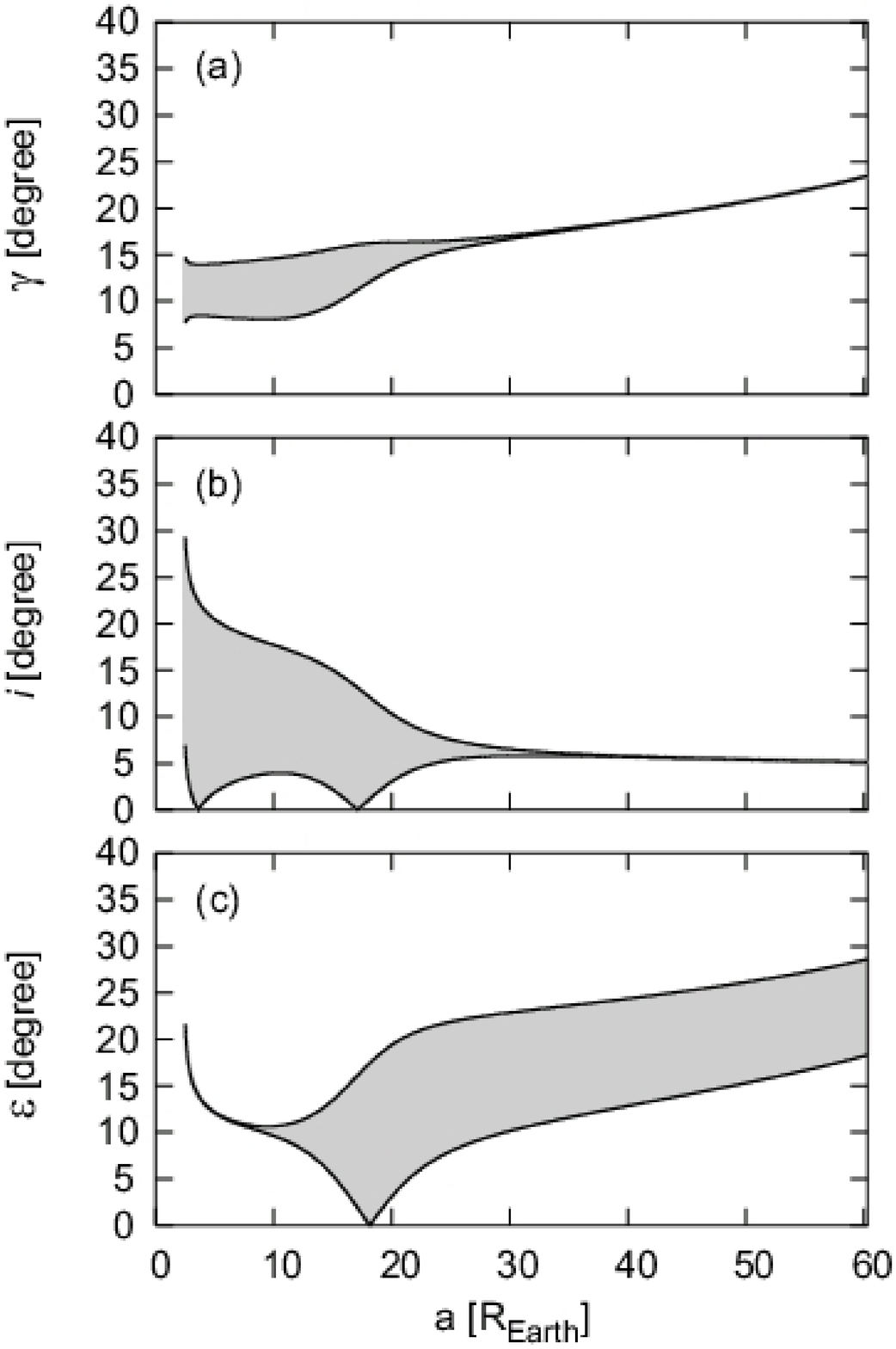}
 \caption{Atobe and Ida}
 \label{fig:EMsys1}
\end{figure}

\begin{figure}
 \centering 
 \includegraphics[width=10cm,clip]{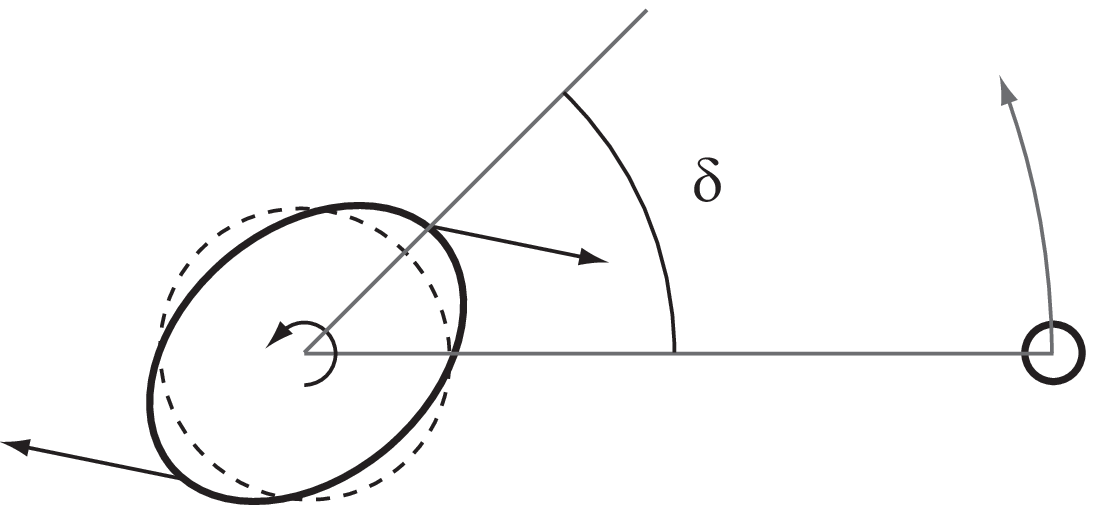}
 \caption{Atobe and Ida}
 \label{fig:config1}
\end{figure}

\begin{figure}
 \centering 
 \includegraphics[width=15cm,clip]{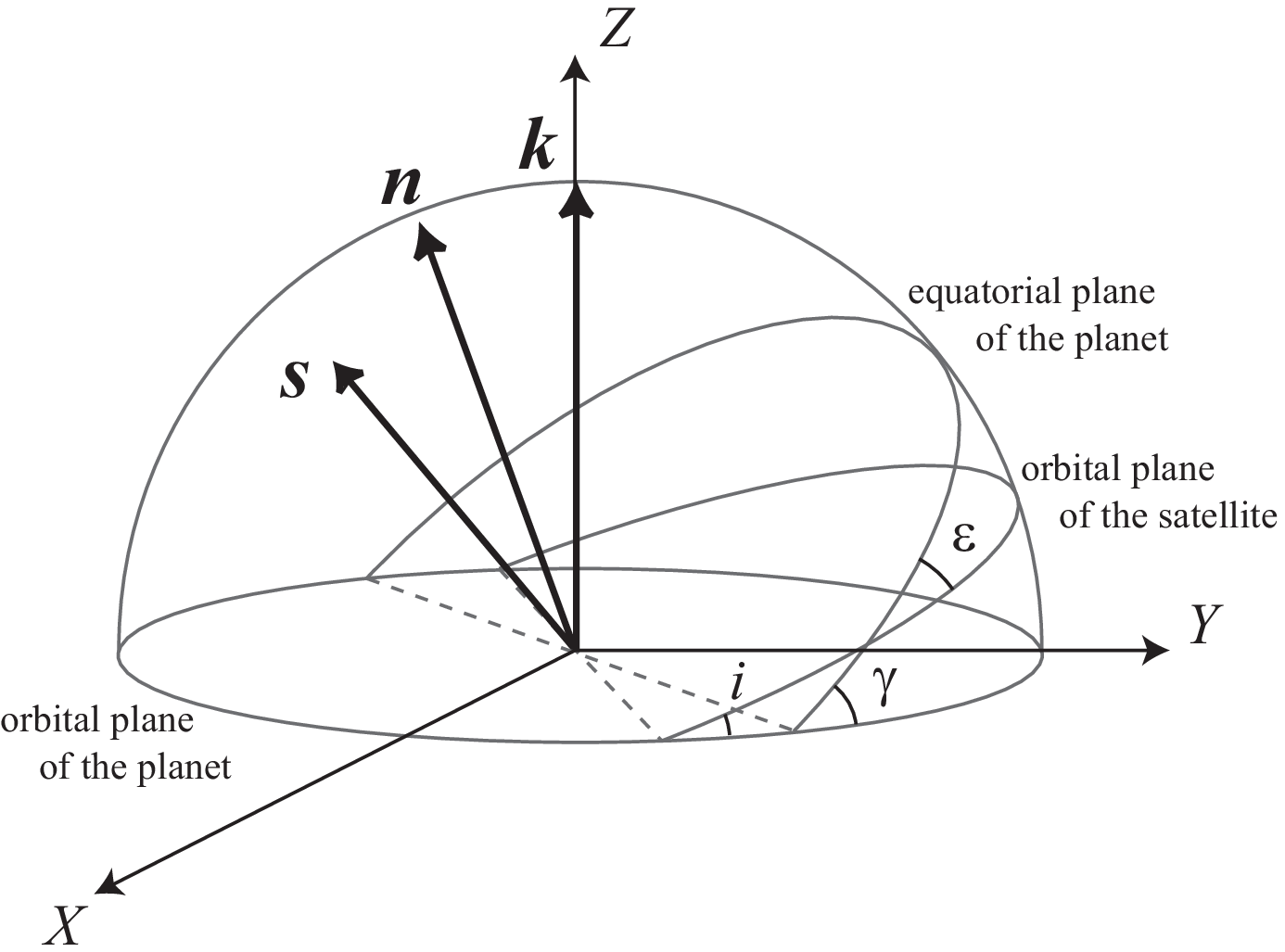}
 \caption{Atobe and Ida}
 \label{fig:config2}
\end{figure}

\begin{figure}
 \centering 
 \includegraphics[width=17cm,clip]{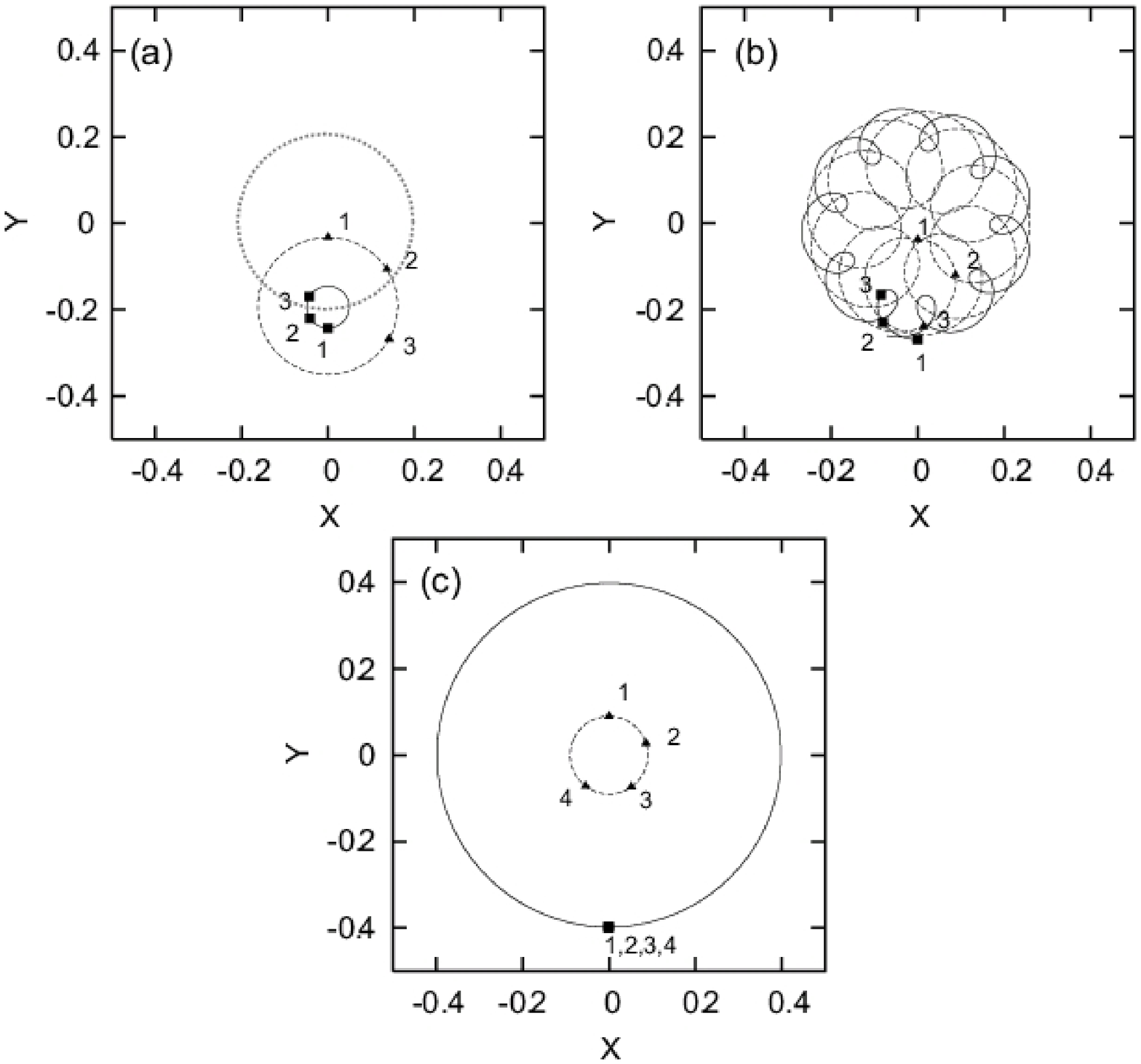}
 \caption{Atobe and Ida}
 \label{fig:EMsys3}
\end{figure}

\begin{figure}
 \centering 
 \includegraphics[width=17cm,clip]{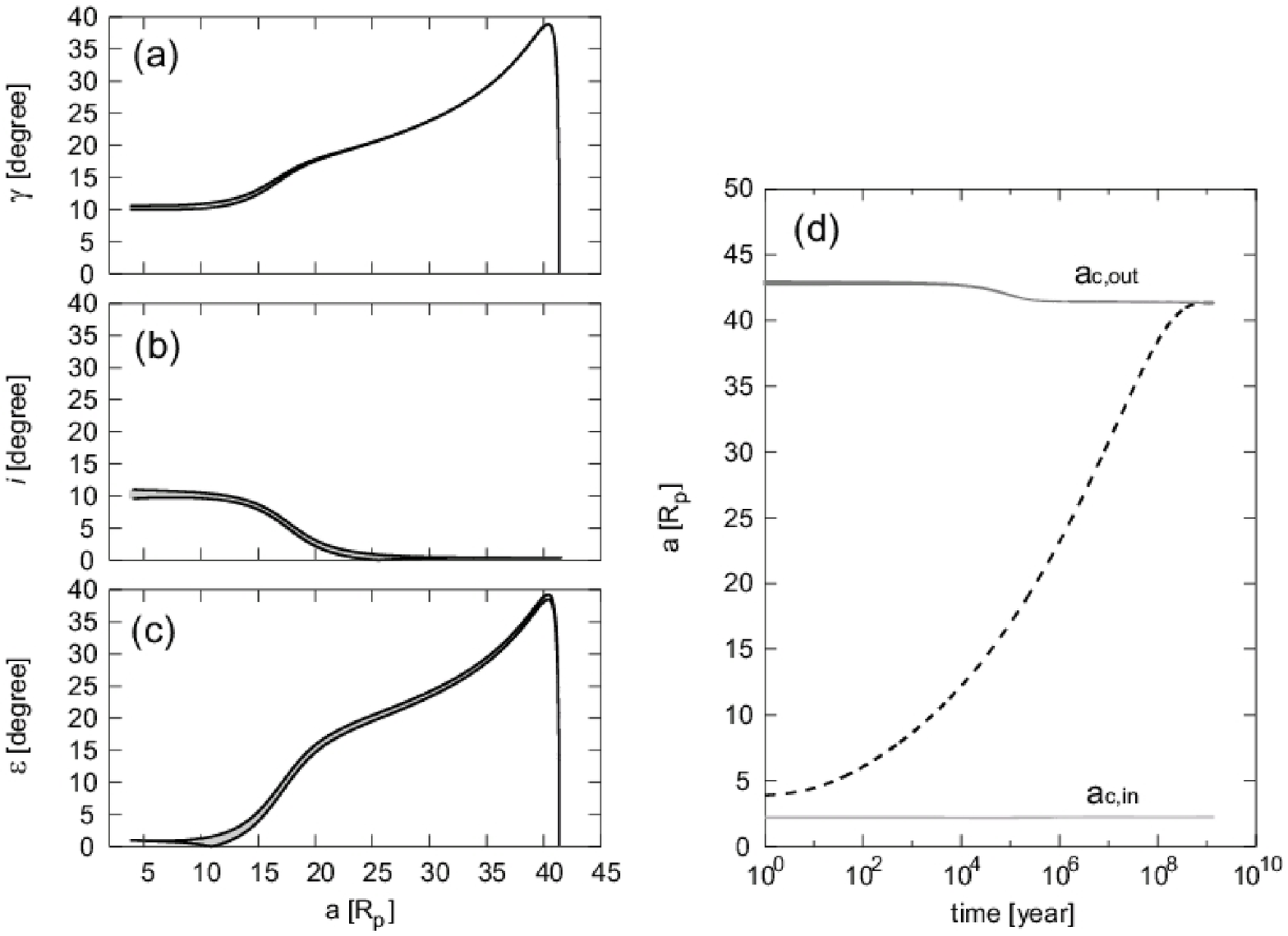}
 \caption{Atobe and Ida}
 \label{fig:fig1}
\end{figure}

\begin{figure}
 \centering 
 \includegraphics[width=17cm,clip]{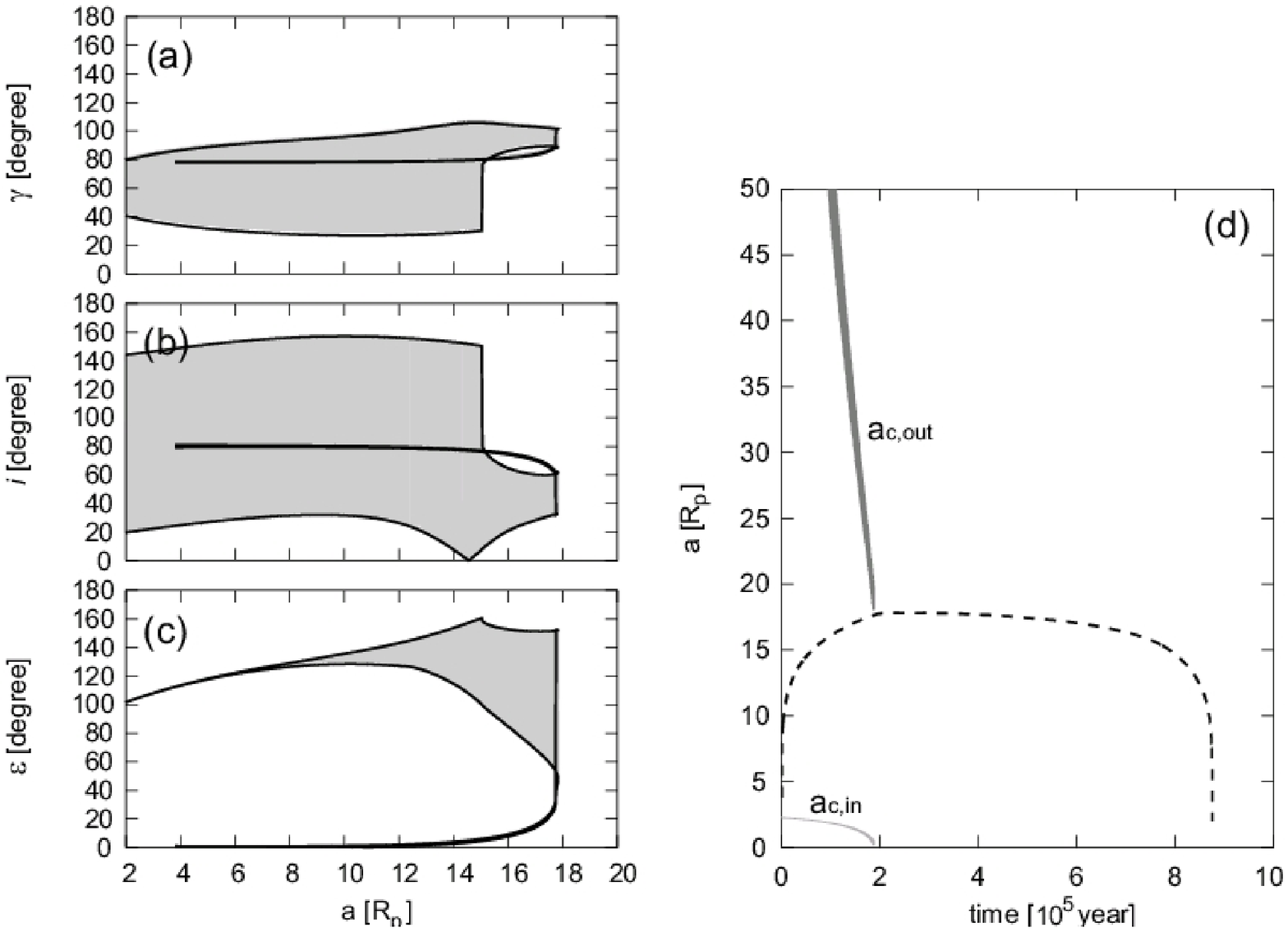}
 \caption{Atobe and Ida}
 \label{fig:fig2}
\end{figure}

\begin{figure}
 \centering 
 \includegraphics[width=10cm,clip]{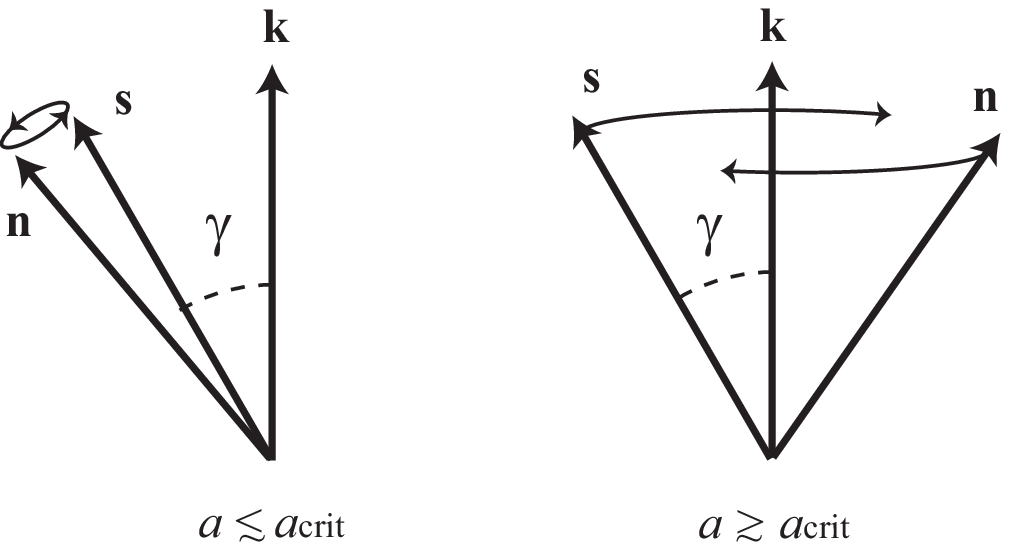}
 \caption{Atobe and Ida}
 \label{fig:config3}
\end{figure}

\begin{figure}
 \centering 
 \includegraphics[width=17cm,clip]{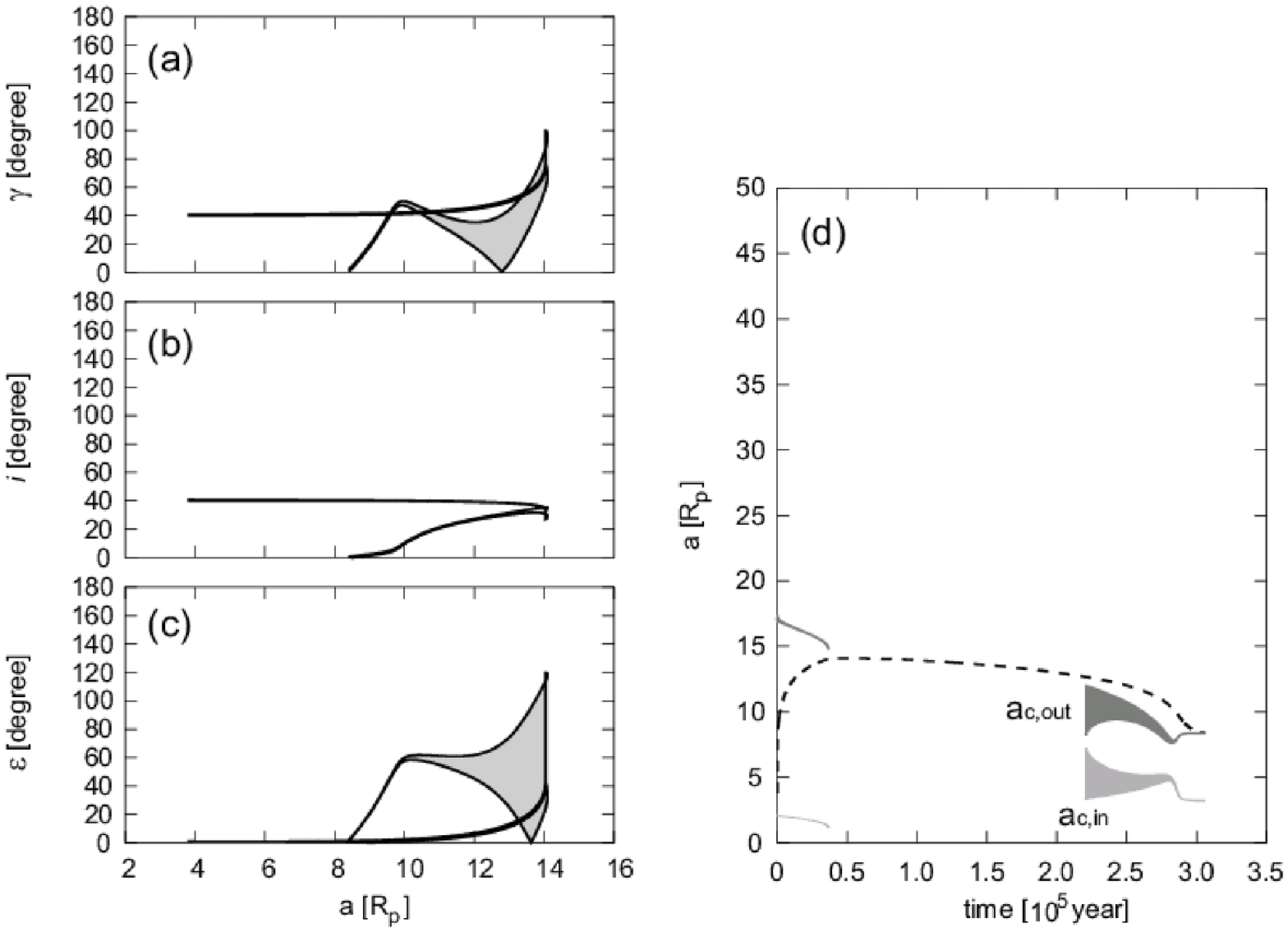}
 \caption{Atobe and Ida}
 \label{fig:fig3}
\end{figure}

\begin{figure}
 \includegraphics[width=15cm,clip]{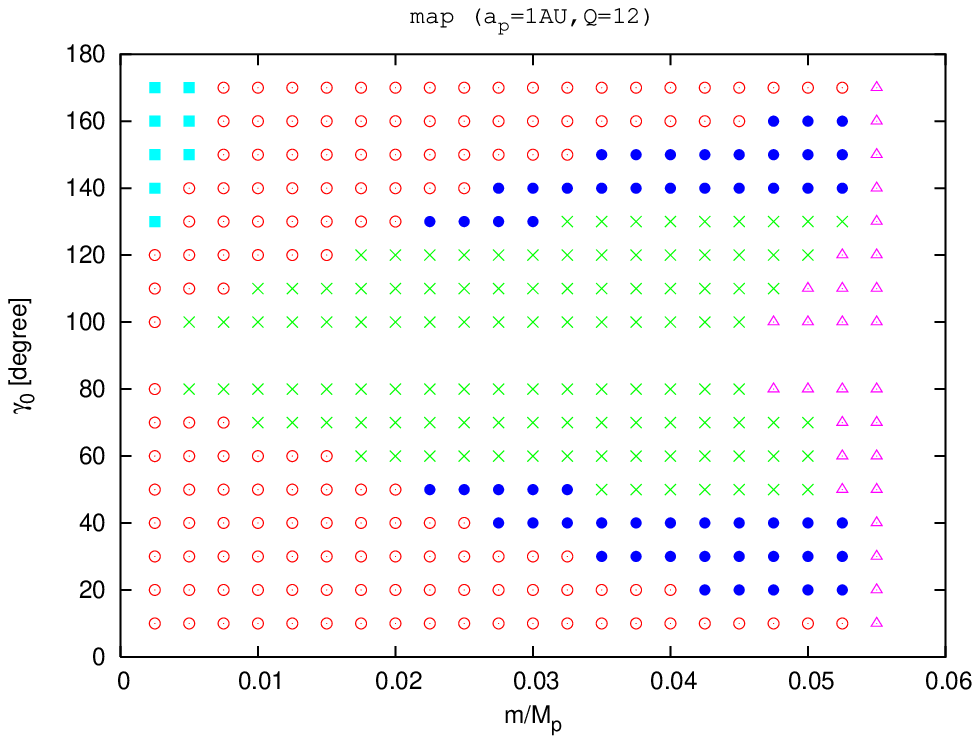}
 \caption{Atobe and Ida}
 \label{fig:numerical}
\end{figure}

\begin{figure}
 \centering 
 \includegraphics[width=13cm,clip]{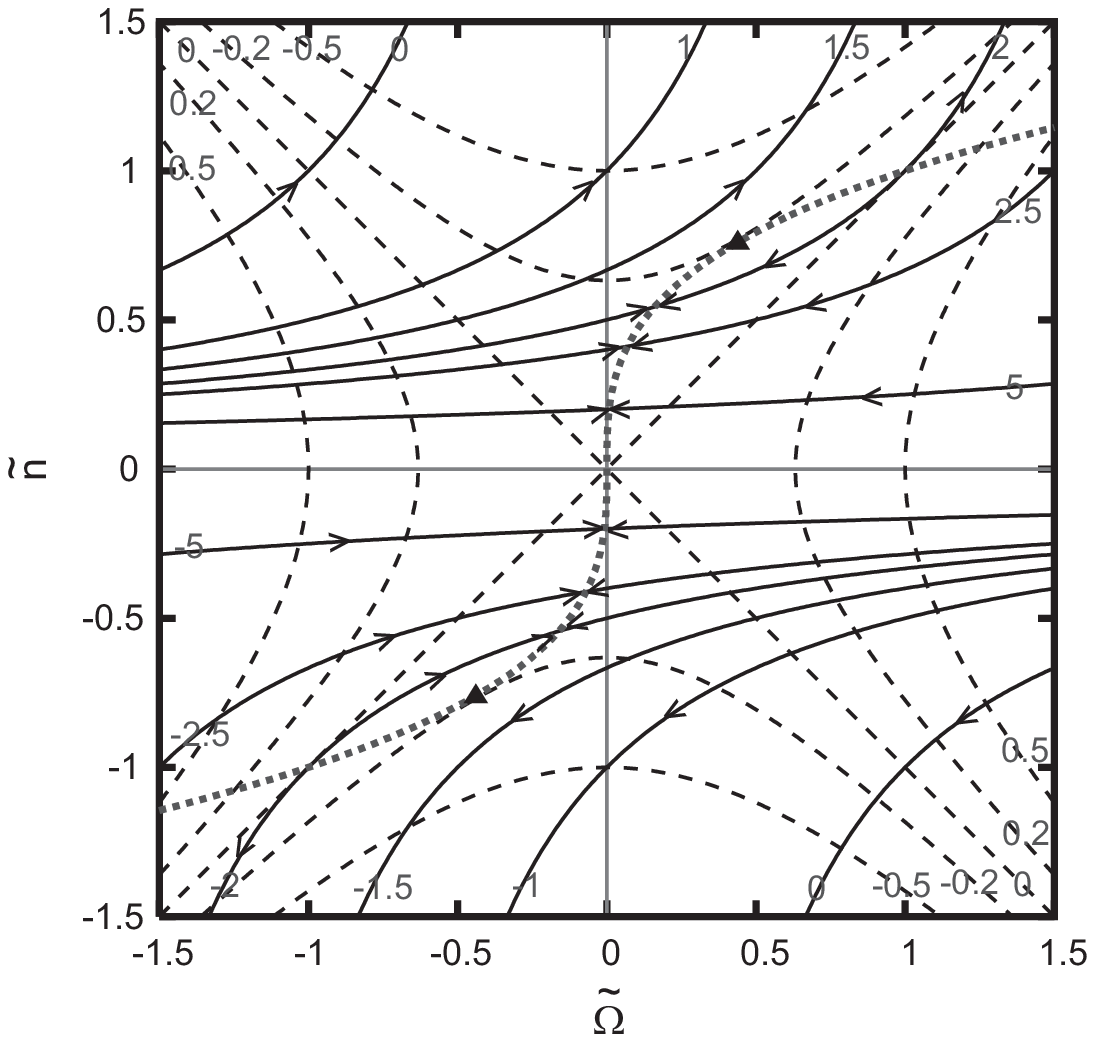}
 \caption{Atobe and Ida}
 \label{fig:cont1}
\end{figure}

\begin{figure}
 \centering 
 \includegraphics[width=10cm,clip]{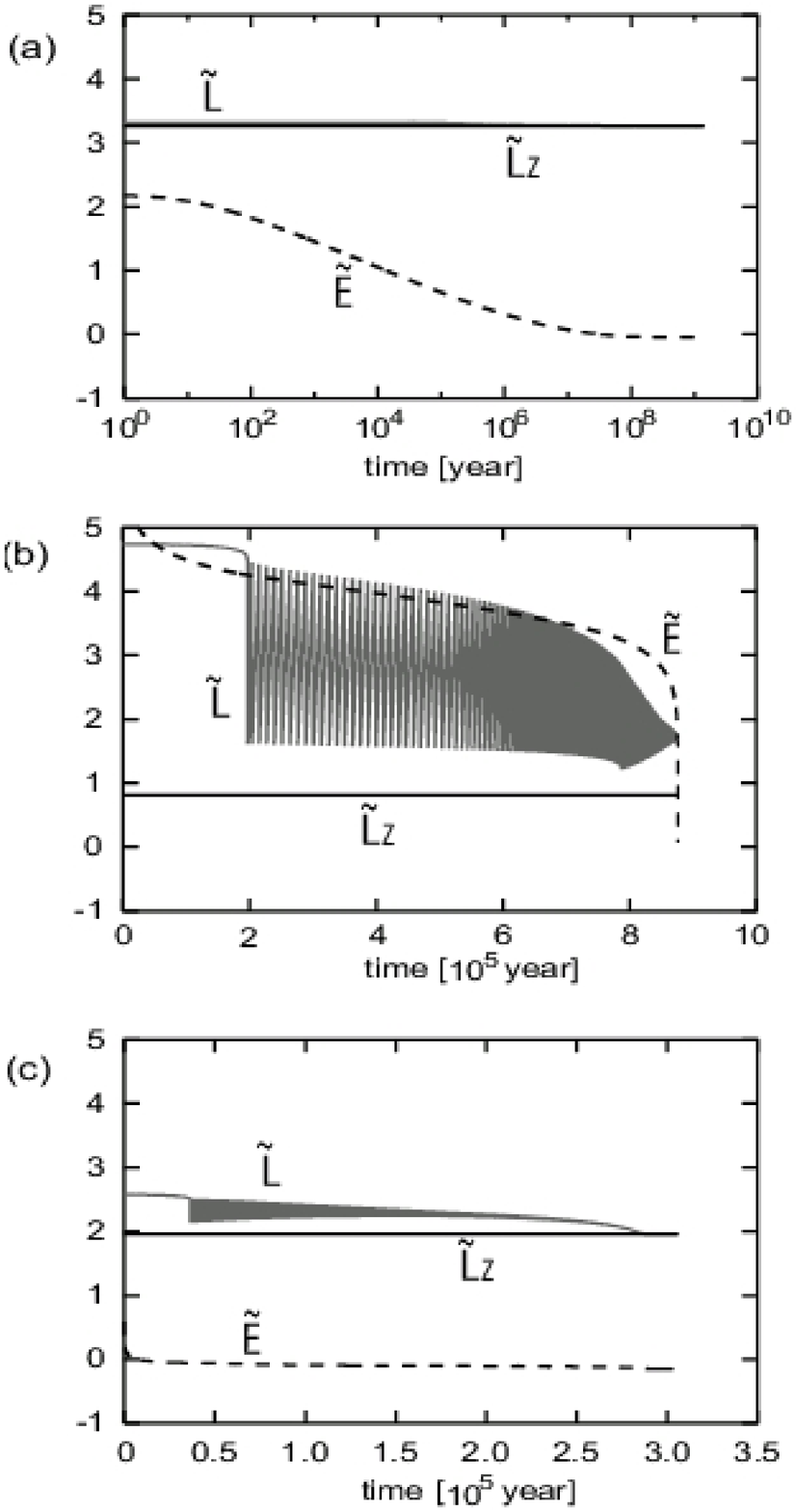}
 \caption{Atobe and Ida}
 \label{fig:L}
\end{figure}

\begin{figure}
 \centering 
 \includegraphics[width=13cm,clip]{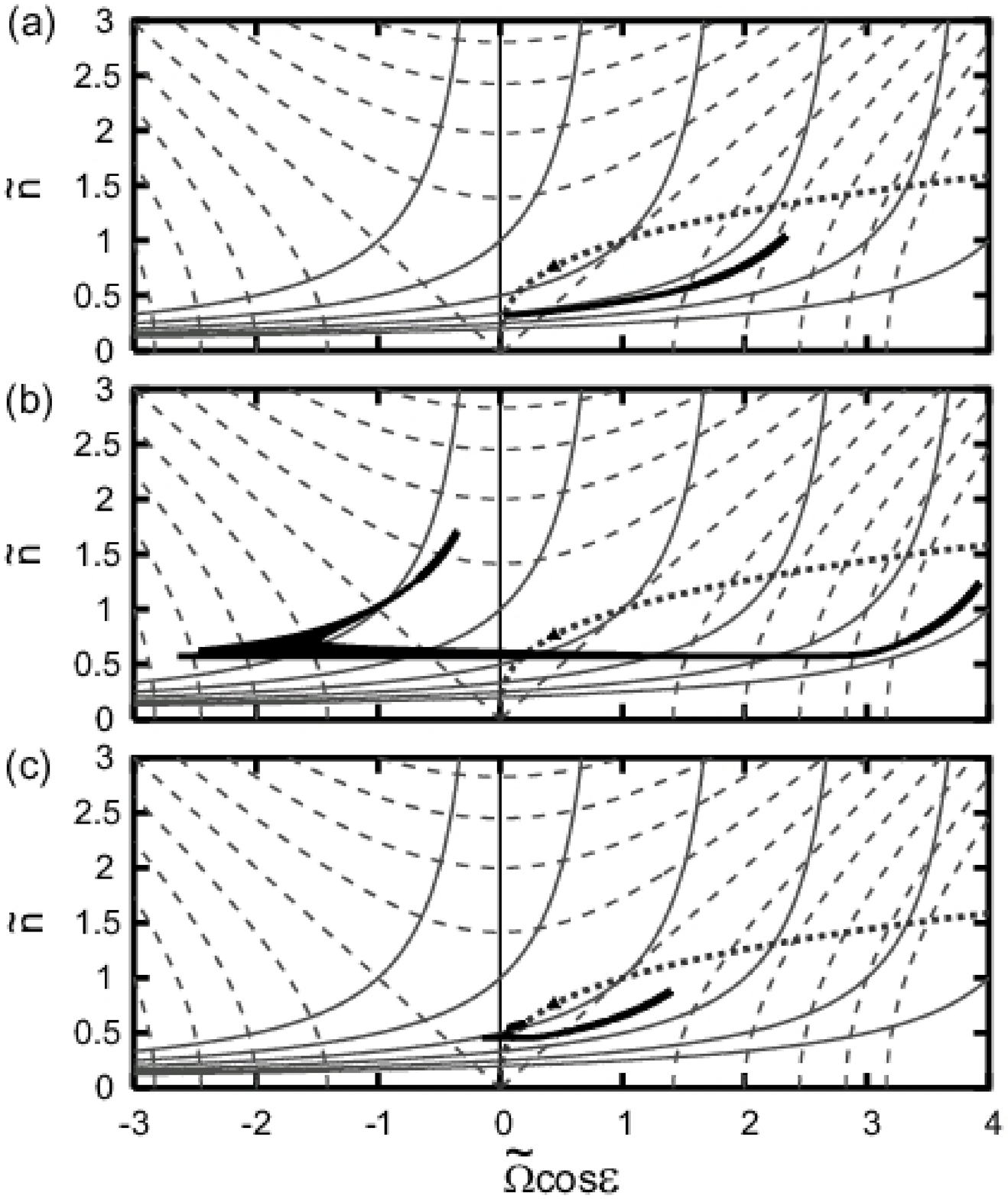}
 \caption{Atobe and Ida}
 \label{fig:cont2}
\end{figure}

\begin{figure}
 \centering 
 \includegraphics[width=15cm,clip]{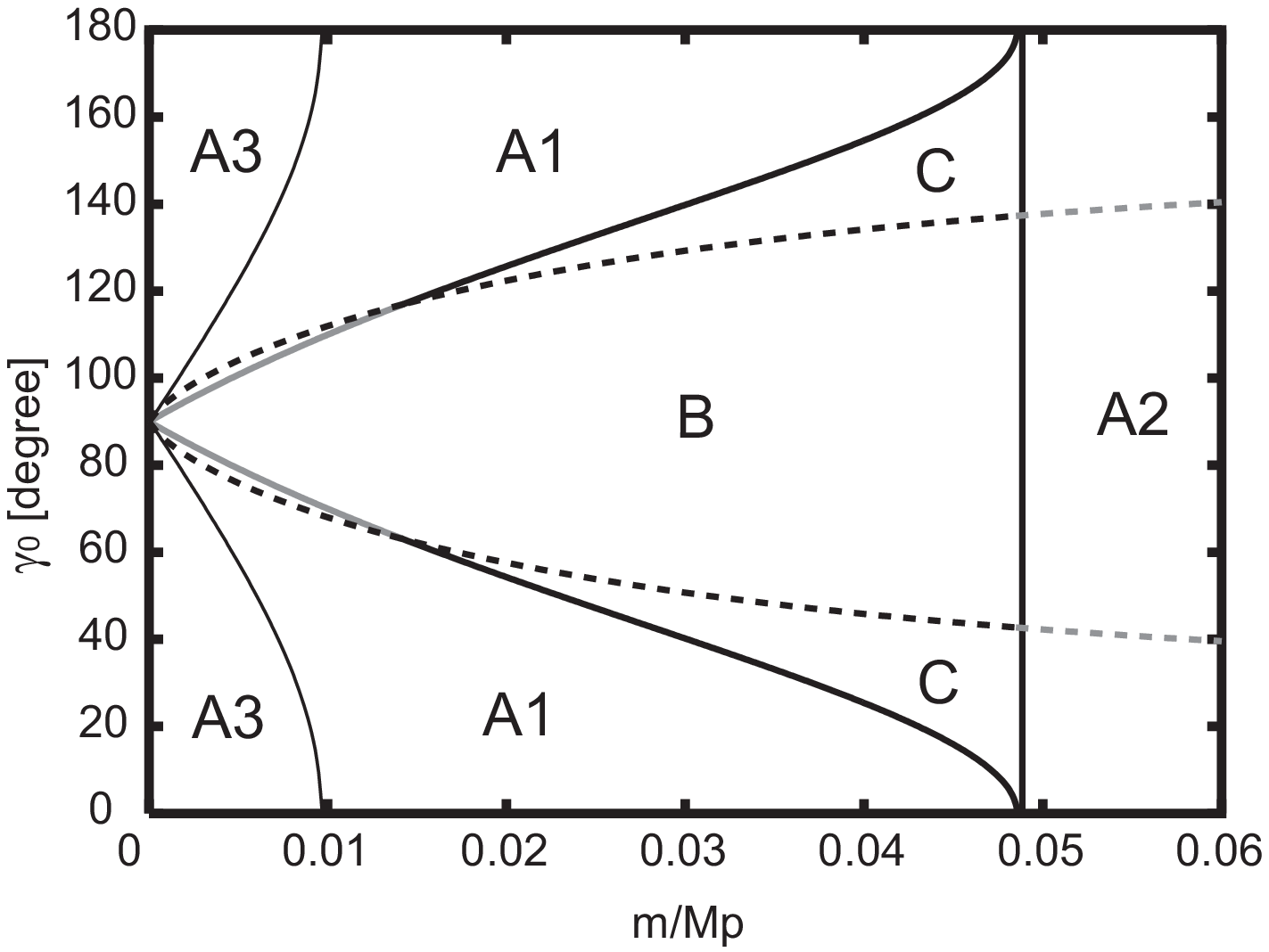}
 \caption{Atobe and Ida}
 \label{fig:a1Q12map1}
\end{figure}

\begin{figure}
 \includegraphics[width=15cm,clip]{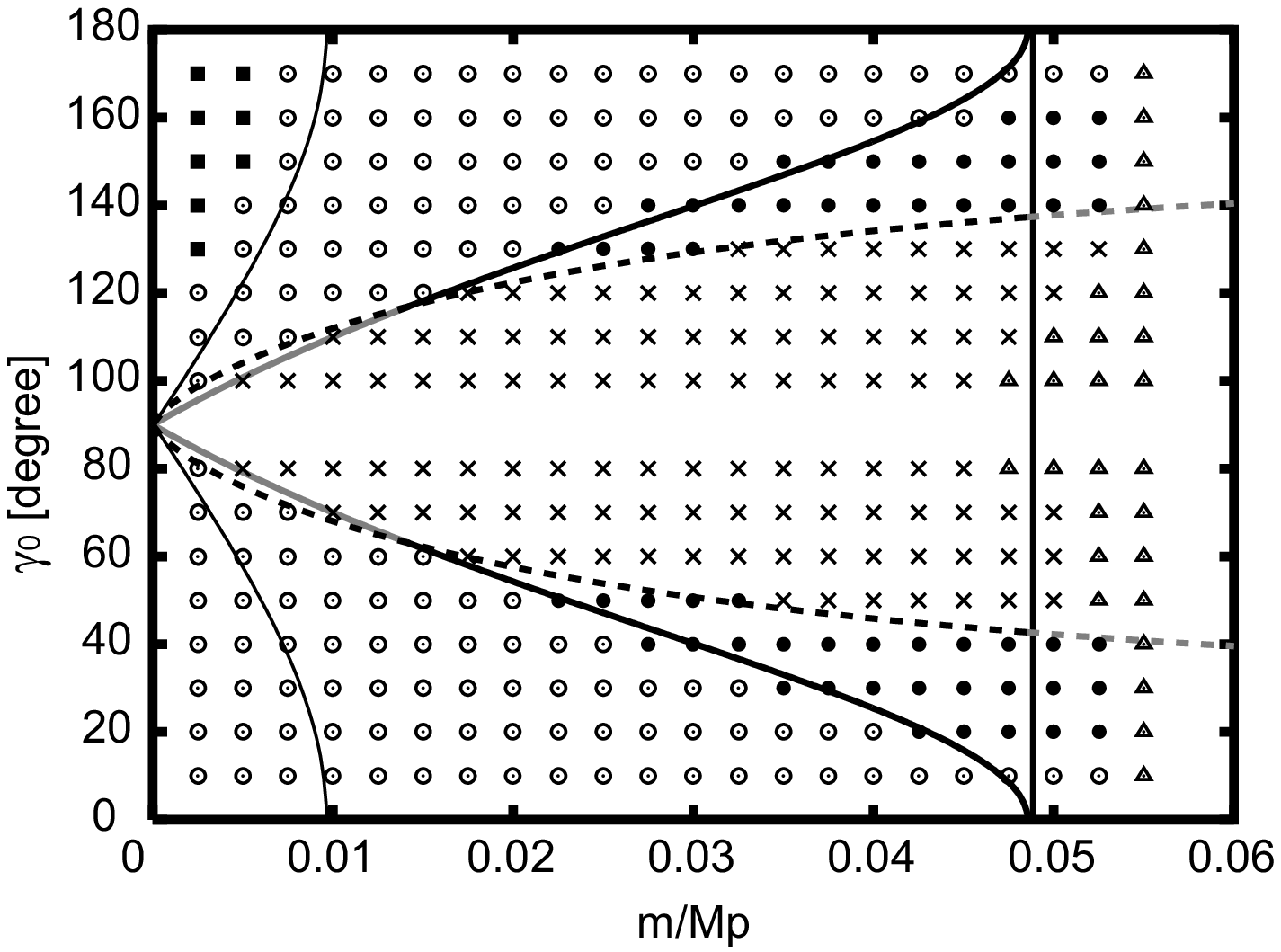}
 \caption{Atobe and Ida}
 \label{fig:a1Q12map1_2}
\end{figure}

\begin{figure}
 \includegraphics[width=15cm,clip]{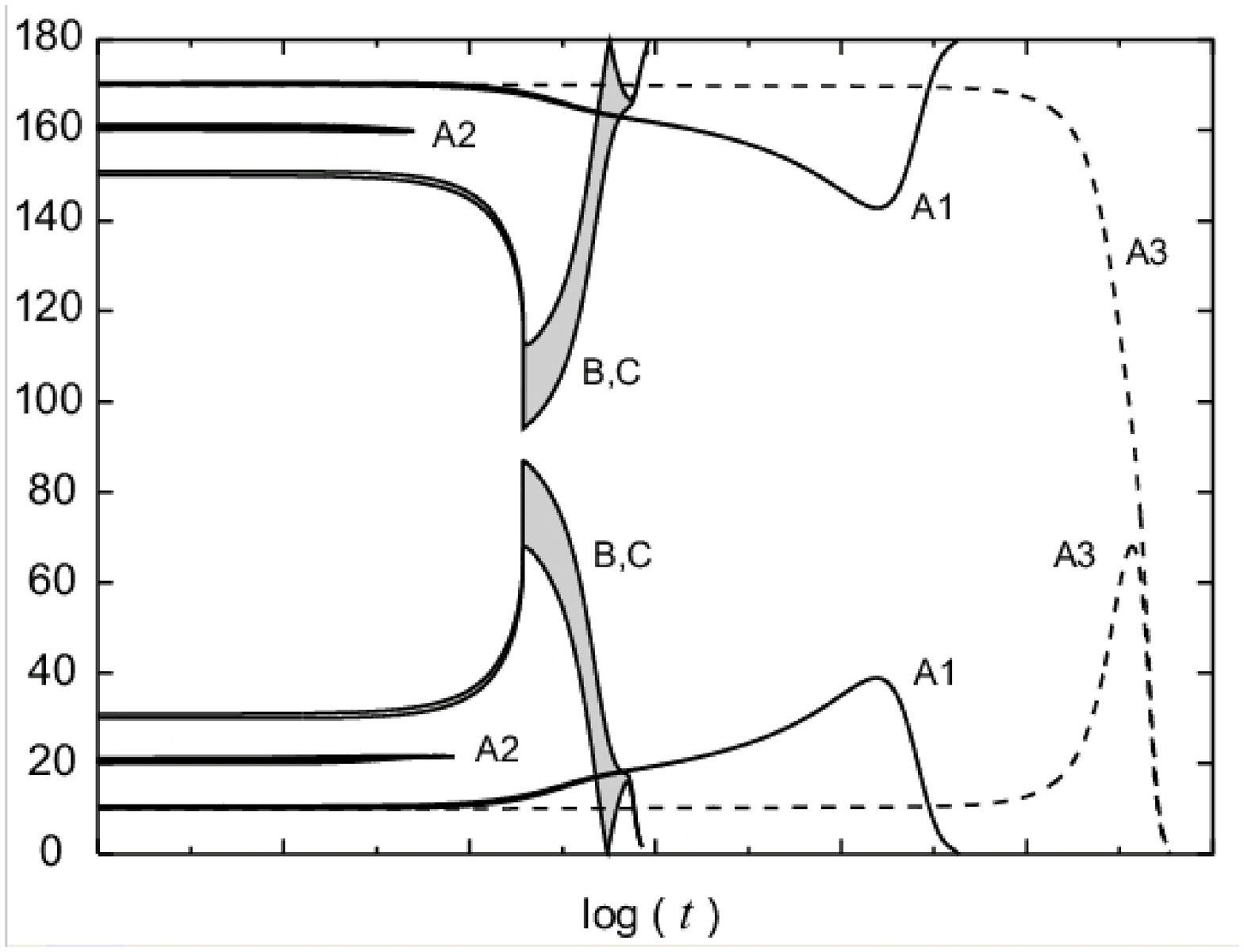}
 \caption{Atobe and Ida}
 \label{fig:obliquity}
\end{figure}

\begin{description}
\item[Figure \ref{fig:EMsys1}]
The tidal evolution of the Earth-Moon system: 
(a) the obliquity of the Earth to the ecliptic $\gamma$, 
(b) the inclination of the Moon's orbit to the ecliptic $i$, 
(c) that to the Earth's equator $\epsilon$. 
The ranges of oscillation during precession are indicated by 
gray patches. 

\item[Figure \ref{fig:config1}] 
Tidal deformation and phase lag are illustrated. 
The planet (represented by 
the circle in dashed line) is distorted (oval in thick solid 
line) by the gravitational force of the external body (the small 
circle in thick solid line). The 
tidal friction causes a phase shift ($\delta$) in the 
response of a planet. The case in which the planetary spin 
is faster than the external body's orbital motion is shown.

\item[Figure \ref{fig:config2}] 
The geometry of the orbital plane of the planet,
its equatorial plane, 
and the orbital plane of the satellite is shown.
Their unit normal vectors are 
$\mathbf{s}$, $\mathbf{k}$, and $\mathbf{n}$.
The obliquity $\gamma$, the inclinations $i$ and 
$\epsilon$ are indicated.

\item[Figure \ref{fig:EMsys3}] 
The trajectories of the tips of $\mathbf{s}$ and $\mathbf{n}$ 
projected onto the orbital plane ($X$-$Y$) 
of the planet in the case of the Earth-Moon system. 
(a) $a \simeq 5R_{\oplus}$, 
(b) $a \simeq 15R_{\oplus}$, 
(c) $a \simeq 60R_{\oplus}$ (the present configuration). 
The solid and dashed lines show 
trajectories of $\mathbf{s}$ and $\mathbf{n}$, respectively. 
Filled squares and triangles indicate the positions 
of $\mathbf{s}$ and $\mathbf{n}$. 
They move on the trajectories in labeled number order. 

\item[Figure \ref{fig:fig1}] 
The numerical calculated tidal evolution of the system with 
$m=0.02M_{\rm p}$, $\gamma_0=10^{\circ}$: 
(a) the obliquity $\gamma$, 
(b) the inclination of the satellite orbit to the planetary orbit $i$, 
(c) that to the planet's equator $\epsilon$, and 
(d) the semi-major axis of the satellite $a$ (dashed line), 
the inner co-rotation radius $a_{\rm c,in}$ (gray solid line) and 
the outer one $a_{\rm c,out}$ (black solid line) calculated 
by $L_{Z0} = L_{Z{\rm c}}$ (Eqs.~\ref{eq:L_Z0} and 
\ref{eq:L_Zc}).
In the panels (a), (b), and (c), the ranges of oscillation during 
precession are indicated by gray patches. 

\item[Figure \ref{fig:fig2}] 
Same as Fig.~\ref{fig:fig1}, but $m=0.01M_{\rm p}$, 
$\gamma_0=80^{\circ}$.

\item[Figure \ref{fig:config3}] 
Schematic illustration of the precession motion of 
$\mathbf{s}$ and $\mathbf{n}$. 
When $a \lesssim a_{\rm crit}$, $\mathbf{s}$ and $\mathbf{n}$ 
precess about their total vector (type I precession). 
During the precession, the mutual 
inclination ($\epsilon$) between $\mathbf{s}$ and $\mathbf{n}$ 
is almost constant. When $a \gtrsim a_{\rm crit}$, 
$\mathbf{s}$ and $\mathbf{n}$ precess about $\mathbf{k}$, 
respectively (type III precession). 
During the precession, $\epsilon$ oscillates
from 0$^{\circ}$ to $\sim 2\gamma$.

\item[Figure \ref{fig:fig3}] 
Same as Fig.~\ref{fig:fig1}, but $m=0.04M_{\rm p}$, 
$\gamma_0=40^{\circ}$.

\item[Figure \ref{fig:numerical}] 
The dependence of tidal evolution on initial conditions
(initial obliquity $\gamma_0$ and satellite-planet mass ratio
$m/M_{\rm p}$) found by the numerical integration. 
Type A1 evolution with 
$\gamma \rightarrow 0^{\circ}$ 
(for $\gamma_0 < 90^{\circ}$) or $\gamma \rightarrow 180^{\circ}$ 
(for $\gamma_0 > 90^{\circ}$) is plotted by open circles. 
A3 evolution which shows $\gamma \rightarrow 0^{\circ}$ with 
$\gamma_0 > 90^{\circ}$ is plotted by filled squares. 
Crosses and filled circles represent B and C evolution,
respectively. 
Triangles represent A2 evolution. 

\item[Figure \ref{fig:cont1}] 
Contours of angular momentum $\tilde{L}$ (solid lines) 
and energy $\tilde{E}$ (dashed lines) in a planar system. 
Planetary spin and satellite's 
orbital periods are equal when $\tilde{\Omega} = \tilde{n}^3$, represented 
by the dotted line. Critical points are shown by filled triangles, 
at which 
all lines of constant-$\tilde{L}$, constant-$\tilde{E}$, and 
synchronism are mutually tangent. Path of tidal evolution in the 
$(\tilde{\Omega},\tilde{n})$-plane are indicated by arrows.

\item[Figure \ref{fig:L}] 
The time evolution of the normalized total angular momentum 
($\tilde{L}$) and its $Z$-component ($\tilde{L_{\rm Z}}$), 
and the total mechanical energy of the planet-satellite system 
($\tilde{E}$), in the cases of 
(a)$m=0.02M_{\rm p}$ and $\gamma_0=10^{\circ}$, 
(b)$m=0.01M_{\rm p}$ and $\gamma_0=80^{\circ}$ and 
(c)$m=0.04M_{\rm p}$ and $\gamma_0=40^{\circ}$. 

\item[Figure \ref{fig:cont2}] 
Path of the tidal evolution of non-planar systems
in the case of 
(a) $m=0.02M_{\rm p}$ and $\gamma_0=10^{\circ}$, 
(b) $m=0.01M_{\rm p}$ and $\gamma_0=80^{\circ}$, 
(c) $m=0.04M_{\rm p}$ and $\gamma_0=40^{\circ}$. 
Thick solid lines are the numerical results.
Solid and dashed lines are contours 
of $\tilde{L}$ lines assuming $\epsilon = 0$ and
$\tilde{E}$. Spin-orbit 
synchronism ($\tilde{\Omega} \cos \epsilon = \tilde{n}^3$) is 
represented by the dotted lines. 

\item[Figure \ref{fig:a1Q12map1}] 
Qualitatively different tidal evolution predicted by the theoretical 
arguments, as a function of $(m/M_{\rm p},\gamma_0)$. 
B and C are separated by $\vert \tilde{L}_Z \vert \simeq \tilde{L}^{\ast}$ 
(thick dashed lines). 
A1 and B are separated by $a_{\rm crit} \simeq a_{\rm c,out}^{\infty}$ 
(thick solid line). 
A2 and B/C are separated by the vertical line which has the 
mass determined by $a_{\rm crit} \simeq a_{\rm c,out}^{\infty}$ 
with $\gamma_0=0^{\circ}$. 
The boundary between A1 and A3 is given by 
$\Delta t_{\rm A3} = \Delta t_{\rm A1}$
(thin solid lines). 

\item[Figure \ref{fig:a1Q12map1_2}] 
Comparison of the analytically 
estimated boundaries with the numerical results in 
Fig.~\ref{fig:numerical}.  For 
the analytical estimate, see text.

\item[Figure \ref{fig:obliquity}] 
Schematic illustration of typical obliquity changes 
in individual evolution types.

\item[Figure \ref{fig:a1Q12map2_2}] 
The timescales of the tidal evolution.
with $Q=12$ and 
$k_2=0.30$, using Eqs.(\ref{Dt_A2}). 
The contours of the timescales are represented by
dotted lines. 
From right to left, contours represent $10^5, 10^6, ..., 10^{12}$ yrs.

\end{description}
%%%%% introduction %%%%%

\clearpage

\clearpage

\clearpage

\clearpage

\clearpage

\clearpage

\clearpage

\clearpage

\clearpage

\clearpage

\clearpage

\clearpage

\clearpage

\clearpage

\clearpage

\begin{figure}
 \centering 
 \includegraphics[width=15cm,clip]{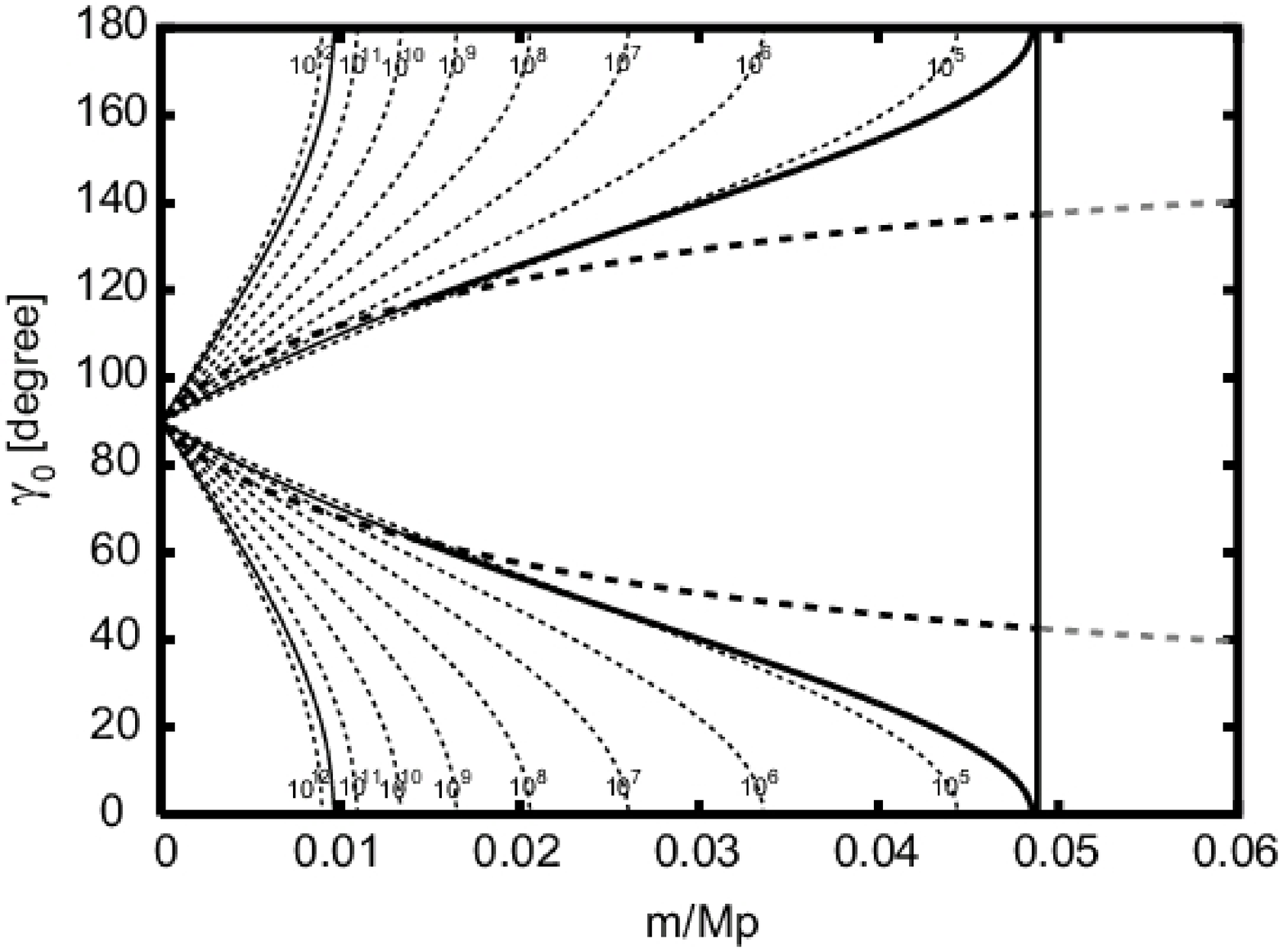}
 \caption{Atobe and Ida}
 \label{fig:a1Q12map2_2}
\end{figure}

\end{document}